\documentclass[12pt]{article}
\usepackage{amsmath}
\usepackage{graphicx}
\usepackage{natbib}
\usepackage{url} 
\usepackage{bbm}
\usepackage{enumerate}
\usepackage{cancel}
\usepackage{multirow}
\usepackage{xcolor}

\newcommand{\bs}{\boldsymbol}
\newcommand{\bth}{\bs{\theta}}

\newcommand{\by}{\bs{y}}

\def\bSig\mathbf{\Sigma}

\newcommand{\prob}{\mathbbm{P}}

\newtheorem{theorem}{Theorem}

\newcommand{\blind}{0}

\begin{document}

\def\spacingset#1{\renewcommand{\baselinestretch}%
{#1}\small\normalsize} \spacingset{1}


\if0\blind
{
  \title{\bf Fast Bayesian Inference for Spatial Mean-Parameterized Conway--Maxwell--Poisson Models}
  \author{Bokgyeong Kang\\
    Department of Statistical Science, Duke University\\
    and \\
    John Hughes 
    \\
    College of Health, Lehigh University\\
    and \\
    Murali Haran\\
    Department of Statistics, The Pennsylvania State University}
  \maketitle
} \fi

\if1\blind
{
  \bigskip
  \bigskip
  \bigskip
  \begin{center}
    {\LARGE\bf Title}
\end{center}
  \medskip
} \fi

\bigskip
\begin{abstract}
Count data with complex features arise in many disciplines, including ecology, agriculture, criminology, medicine, and public health. Zero inflation, spatial dependence, and non-equidispersion are common features in count data. There are two classes of models that allow for these features---the mode-parameterized Conway--Maxwell--Poisson (COMP) distribution and the generalized Poisson model. However both require the use of either constraints on the parameter space or a parameterization that leads to challenges in interpretability. We propose spatial mean-parameterized COMP models that retain the flexibility of these models while resolving the above issues. We use a Bayesian spatial filtering approach in order to efficiently handle high-dimensional spatial data and we use reversible-jump MCMC to automatically choose the basis vectors for spatial filtering. The COMP distribution poses two additional computational challenges---an intractable normalizing function in the likelihood and no closed-form expression for the mean. We propose a fast computational approach that addresses these challenges by, respectively, introducing an efficient auxiliary variable algorithm and pre-computing key approximations for fast likelihood evaluation. We illustrate the application of our methodology to simulated and real datasets, including Texas HPV-cancer data and US vaccine refusal data.
\end{abstract}

\noindent%
\vfill

\newpage
\spacingset{1.5} 






\section{Introduction}
\label{sec:intro}

Count data exhibiting zero inflation, spatial dependence, and non-equidispersion are commonly encountered in many disciplines, including environmental sciences, agriculture, politics, criminology, medicine, public health studies, and manufacturing applications \citep[cf.][]{Ratcliffe2007,Neelon2016,Yang2022}. 
The Conway--Maxwell--Poisson (COMP) distribution \citep{Conway1962} is a two-parameter generalization of the Poisson distribution. It has rate $\lambda$ and dispersion $\nu$ that allows for under- and overdispersion.
The COMP distribution has several appealing properties compared to existing alternatives. For example, the COMP distribution belongs to the exponential and the power series families of distributions, and so offers sufficient statistics and permits elegant derivation of additional properties \citep{Shmueli2005}. 
There are a number of alternative flexible count models but each one has some limitations.
For instance, the popular negative binomial model can only capture overdispersion. The generalized Poisson distribution \citep{Famoye1993} can accommodate only a limited degree of underdispersion. The double Poisson distribution \citep{Efron1986} requires approximation to its normalizing constant which may be highly unreliable for small means \citep{Zou2013}. Finally, the Gamma-count distribution \citep{Winkelmann1995} cannot be parameterized via the mean. The spatial mean-parameterized COMP models we develop do not have the above drawbacks but they pose a number of computational challenges. Hence, a significant contribution of our paper is developing a computationally efficient approach for fitting these models. 

Broader application of the COMP distribution has been thwarted by the fact that there is no closed-form expression for the mean. This makes it difficult to interpret results in regression settings, and renders the COMP regression model incomparable to competing models such as the log-linear Poisson, negative binomial, and generalized Poisson regression models. \citet{Guikema2008} and \citet{Ribeiro2020} approximated the mode and mean, respectively, of the COMP distribution, and reparameterized the distribution accordingly. However, these approximations are accurate only for a restricted parameter space. \citet{Huang2017} parameterized the distribution by introducing a rate function $\lambda(\mu, \nu)$ that depends on the mean $\mu$ and dispersion $\nu$. 
The mean-parameterized COMP (henceforth denoted as COMP$_{\mu}$) distribution has no such restriction of the parameter space. Also, $\mu$ and $\nu$ are orthogonal parameters, which can improve Markov chain mixing compared to alternative parameterizations. For independent outcomes, \citet{Huang2017} used a Newton--Raphson (NR) algorithm to obtain the maximum likelihood estimators (MLEs) of the parameters for COMP$_{\mu}$ regression models. However, it is challenging to measure the uncertainty of the MLE of $\nu$ due to the complicated expression of $\nu$'s standard error. And computation is challenging for spatial models that incorporate high-dimensional spatial effects. \citet{Huang2019} proposed a Bayesian approach to fitting COMP$_{\mu}$ regression models, where $\lambda(\mu, \nu)$ is estimated via the NR algorithm at each iteration of Markov chain Monte Carlo (MCMC). This can be computationally expensive even for spatial datasets of moderate size. 
\citet{Philipson2023} introduced a fast Bayesian method that creates a grid of $\log \mu$ and $\nu$ with a recommended step size of 0.01, finds $\lambda$ at each grid vertex using a root-finding method of the Jenkins--Traub (JT) algorithm, and approximates $\lambda(\mu, \nu)$ by bilinear interpolation at each step of MCMC. This method can considerably reduce computation cost and provide reliable parameter estimates for many examples. However, the number of grid vertices grows polynomially with increasing ranges of $\mu$ and $\nu$. For the Texas HPV-cancer data, for instance, because observations $y_i \in \{0, 1, \dots, 1346\}$, we have to consider at least $[0.01, y_{\max} = 1346] \times [0.01, \hat{\nu} = 1.2]$ for the ranges of $\mu$ and $\nu$, which results in 16,152,000 grid vertices. Finding $\lambda$ at 16,152,000 grid vertices is very time-consuming. Also the JT algorithm is numerically unstable and may fail to find roots when $\mu \geq 200$ (see Supplemental Section A.3 for details).
These challenges motivated our development of a new, flexible model and an efficient computational method. 

In this article we propose new spatial COMP$_{\mu}$ models and several computational methods that address the above mentioned challenges. We accommodate spatial clustering by using the Bayesian spatial filtering (BSF) approach \citep{hughes2017spatial}, which models spatial structure as a linear combination of spatially patterned basis functions. This method can improve Markov chain mixing by reducing the dimension of and correlation among the spatial effects. We choose important basis vectors using a reversible jump MCMC \citep[RJMCMC; ][]{Green1995,Godsill2012} algorithm. Our chief computational challenge stems from the lack of a closed-form expression for the COMP$_{\mu}$ mean. 
We propose a spline approximation to $\lambda(\mu, \nu)$, which allows for reliable approximation and fast computation for a wide range of mean.
Bayesian inference for the class of COMP models is challenging due to an intractable normalizing function in the likelihood. We overcome this problem by using an exchange algorithm \citep{murray2006}, which introduces auxiliary variables so that the intractable terms cancel out in the acceptance ratio. The exchange algorithm may be impractical when data exhibit zero inflation and severe overdispersion. We introduce a proposal distribution that permits the algorithm to perform better for zero-inflated models. Positive spatial dependence can cause extra underdispersion. Applying our methodology to challenging simulated and real datasets shows that our method can effectively distinguish spatial dependence from dispersion.

The remainder of this article is organized as follows. In Section~\ref{sec:model} we specify our spatial COMP$_{\mu}$ and ZICOMP$_{\mu}$ regression models. In Section~\ref{sec:algorithm} we propose our fast Bayesian approach to inference and provide implementation details. Theoretical justification for our approach is also provided. In Section~\ref{sec:sim} we discuss simulation experiments for our proposed models and computational approaches. In Section~\ref{sec:real} we analyze Texas HPV-cancer data and US vaccine refusal data. We conclude in Section~\ref{sec:discuss}.
The source code can be download from the journal website or the following repository (\url{https://github.com/bokgyeong/meanCOMP}).

\section{Spatial mean-parameterized COMP models}
\label{sec:model}

In this section we describe the COMP distribution and introduce our spatial COMP$_{\mu}$ and ZICOMP$_{\mu}$ regression models.

\subsection{COMP distribution}
\label{subsec:comp}

The COMP distribution is a two-parameter generalization of the Poisson distribution. 
A count variable $Y$ is said to follow the COMP($\lambda$, $\nu$) distribution if $Y$'s probability mass function (pmf) is given by
\begin{align*}
    \prob(Y = y) &= \frac{1}{c(\lambda, \nu)} \frac{\lambda^y}{(y!)^{\nu}},
\end{align*}
where $\lambda > 0$ is a rate parameter, $\nu \geq 0$ is a dispersion parameter, and $c(\lambda, \nu)$ $=$ $ \sum_{z=0}^{\infty}$ $\lambda^z/(z!)^{\nu}$ is the normalizing function. The COMP distribution possesses several attractive features. First, it allows for underdispersion (variance less than the mean) when $\nu > 1$, equidispersion (variance equals the mean) when $\nu = 1$, and overdispersion (variance greater than the mean) when $0 \leq \nu <1$. Second, the COMP distribution has three classical count distributions as special cases: Poisson ($\nu = 1$), geometric ($\nu = 0$, $\lambda < 1$, and success probability $1-\lambda$), and Bernoulli ($\nu = \infty$ and success probability $\lambda/(1+\lambda)$). Finally, the COMP distribution belongs not only to the exponential family but also to the power series family of distributions, which provides sufficient statistics and permits elegant derivation of other properties \citep{Shmueli2005}.

\citet{Huang2017} reparameterized the COMP distribution in terms of the mean $\mu$. A variable $Y$ is said to follow the COMP$_{\mu}$($\mu$, $\nu$) distribution if $Y$'s pmf is given by
\begin{align*}
    \prob(Y = y) &= \frac{1}{c\left[\lambda(\mu, \nu), \nu\right]} \frac{\lambda(\mu, \nu)^y}{(y!)^{\nu}},
\end{align*}
where the rate $\lambda(\mu, \nu)$ is a function of $\mu$ and $\nu$, which can be found by solving 
\begin{align}
    \sum_{z=0}^{\infty} (z - \mu) \frac{\lambda^{z}}{(z!)^{\nu}} = 0 \label{eq:lamNR}
\end{align}
with respect to $\lambda$.
In addition to all of the standard COMP distribution's key features, the COMP$_{\mu}$ has the appealing property that $\mu$ and $\nu$ are orthogonal. This parameter orthogonality is particularly useful in regression settings. Suppose that $y_i \sim \text{COMP}_{\mu}(\mu_i, \nu)$ with $\log(\mu_i) = \textbf{x}^{\top}_i \bs{\beta}$ for $i = 1, \dots, n$, where $\textbf{x}_i$ is a $p \times 1$ vector of covariates for the $i$th observation, and $\bs{\beta}$ is a vector of regression coefficients. Let $\hat{\bs{\beta}}_{\text{MLE}}$ and $\hat{\nu}_{\text{MLE}}$ be the MLEs of $\bs{\beta}$ and $\nu$, respectively. By the orthogonality of $\bs{\beta}$ and $\nu$, $\hat{\bs{\beta}}_{\text{MLE}}$ and $\hat{\nu}_{\text{MLE}}$ are asymptotically independent \citep{Huang2017}.
R package \texttt{mpcmp} \citep{fung2022mpcmp} provides $\hat{\bs{\beta}}_{\text{MLE}}$ and $\hat{\nu}_{\text{MLE}}$, and the standard error of $\hat{\bs{\beta}}_{\text{MLE}}$. However, it is difficult to quantify the uncertainty of $\hat{\nu}_{\text{MLE}}$ because the corresponding standard error takes a complicated form. Additionally, maximum likelihood estimation can be computationally challenging for spatial models. This motivates our development of a fast Bayesian method for spatial COMP$_{\mu}$ models, where the orthogonality of $\mu$ and $\nu$ can speed mixing of Markov chains.


\subsection{Spatial COMP$_{\mu}$ regression model}
\label{subsec:compReg}

When data exhibit spatial dependence, one might consider the following spatial generalized linear mixed model (SGLMM)
\begin{align*}
    y_i &\sim \text{COMP}_{\mu}(\mu_i, \nu)\\
    \log(\mu_i) &= \textbf{x}^{\top}_i \bs{\beta} + u_i \quad (i = 1,\dots,n),
\end{align*}
where $u_i$ is a spatial random effect for location $i$. For areal data the conditional autoregressive (CAR) model \citep{bym} is a traditional choice of the prior distribution for the spatial effects:
\begin{align}
    \bs{u} = (u_1, \dots, u_n)^\top \mid \tau &\sim \textup{Normal}_{n} \left( \textbf{0}, \textbf{Q}^{-1} / \tau \right), \label{eq:icarW}
\end{align}
where $\tau$ is a conditional precision parameter and $\textbf{Q} = \textup{diag}(\textbf{A} \textbf{1}) - \rho \textbf{A}$ is a precision matrix, where $\textbf{A}$ is the adjacency matrix of the underlying graph, $\textbf{1}$ is the conformable vector of 1's, and $\rho \in [0,1)$ indicates the strength of spatial correlation ($\rho$ = 0 implies spatial independence while $\rho$ near 1 implies strong spatial correlation). Note that $\textbf{Q}$ intuitively accounts for both dependencies ($u_{i}$ and $u_{j}$, $i \neq j$, are independent given their neighbors if and only if $(\textbf{Q})_{ij} = (\textbf{Q})_{ji} = 0$ if and only if areal units $i$ and $j$ are not adjacent) and prior uncertainty (uncertainty about $u_{i}$ is inversely proportional to the number of neighbors of areal unit $i$: $(\textbf{Q})_{ii} = (\textbf{A})_{i} \textbf{1}$ where $(\textbf{A})_{i}$ denotes the $i$th row of $\textbf{A}$). MCMC is a standard approach for handling these spatial effects. For large $n$, however, MCMC is computationally burdensome owing to costly ($\mathcal{O}(n^3)$) evaluations of $n$-dimensional multivariate normal likelihood functions at each iteration. Additionally, strong spatial correlation can lead to slow mixing Markov chains \citep{haran2011gaussian}.

To reduce the dimension of and the cross-correlations among the spatial effects, we can express the spatial effects as linear combinations of well-chosen basis vectors, which produces linear predictor $\textbf{X}\bs{\beta} + \textbf{B}\bs{\delta}$, where $\textbf{B}$ is a basis matrix the columns of which are $q\ll n$ basis vectors, and $\bs{\delta}$ is a $q \times 1$ vector of basis coefficients. There are two common choices of $\textbf{B}$: (1) in restricted spatial regression (RSR), the spatial basis vectors are constrained to be orthogonal to the fixed-effects predictors \citep[cf.][]{Reich2006,Hughes2013}; and (2) in Bayesian spatial filtering (BSF), the spatial basis vectors are constrained to be orthogonal to the intercept \citep{hughes2017spatial}. RSR models have been found to yield low coverage rates for the regression coefficients \citep{Hanks2015,Khan2020,Zimmerman2021}. Our simulation experiments showed that the BSF parameterization does not adversely impact inference for the regression coefficients, and so we adopt the BSF approach. Specifically, the BSF basis matrix $\textbf{B}$ comprises selected eigenvectors of the Moran operator $M=(\textbf{I} - \textbf{1}\textbf{1}^\top / n) \textbf{A} (\textbf{I} - \textbf{1}\textbf{1}^\top / n)$, where $\textbf{I}$ is the $n$ $\times$ $n$ identity matrix. The eigenvectors of $M$ comprise all possible mutually distinct patterns that can arise on the graph and hence can be used to model a spatial random field. The positive (negative) eigenvalues of $M$ correspond to varying degrees of attractive (repulsive) spatial dependence. An eigenvector's pattern has finer scale with decreasing magnitude of the corresponding eigenvalue. In other words, \textbf{B} can accommodate spatial structure at multiple scales. We henceforth assume that \textbf{B} comprises the first $q \ll n$ basis vectors since we expect neighboring observations to be similar (i.e., we discard all of the repulsive patterns). The prior distribution for the basis coefficients is given by $\bs{\delta} \mid \tau \sim \textup{Normal}_{q} \left( \textbf{0}, \textbf{Q}_{\text{B}}^{-1} / \tau \right)$, where $\textbf{Q}_{\text{B}} = \textbf{B}^{\top} \textbf{Q} \textbf{B}$. 
\subsection{Spatial ZICOMP$_{\mu}$ regression model}
\label{subsec:zicompReg}

Zero-inflated models \citep{Lambert1992,greene1994zinb} are mixtures of a zero-degenerate distribution and a count distribution. The zero-degenerate distribution accounts for excess zeros. The count distribution describes the remaining zeros and positive counts. Here the word ``excess'' refers to the fact that there are more zeros in the data than expected under a standard count model. Zero-inflated models separate data into two classes by construction. Structural zeros correspond to subjects who are not at risk for some event and thus have no chance of a positive count. Chance or at-risk observations correspond to subjects who are at risk for the event. Zeros in the chance class apply to subjects who are at risk for an event but nevertheless have 0 as their observed response.

Our ZICOMP$_{\mu}$ distribution has pmf
\begin{align*}
    \prob(Y_{it} = y_{it}) &= (1-\pi_{it}) \mathbbm{1}_{\{w_{it} = 0, y_{it} = 0\}} + \pi_{it} p(y_{it}; \mu_{it}, \nu) \mathbbm{1}_{\{w_{it} = 1\}} \: (i=1,\dots,n) \: (t = 1,\dots,T),
\end{align*}
where $y_{it}$ denotes the observed response in location $i$ at time $t$, $\mathbbm{1}_{\{ \cdot \}}$ is the indicator function, $\pi_{it}$ is the probability of $y_{it}$ being in the chance class, $w_{it}$ is a latent chance indicator variable that follows a Bernoulli distribution with mean $\pi_{it}$, and $p(\cdot\, ; \mu_{it}, \nu)$ is the COMP$_{\mu}$($\mu_{it}, \nu$) pmf. For areal data we can model the chance probability $\pi_{it}$ and the mean $\mu_{it}$ as follows:
\begin{align*}
    \text{logit}(\pi_{it}) &= \textbf{x}_{it}^\top \bs{\beta}_1 + \textbf{b}^\top_i \bs{\gamma}\\
    \log(\mu_{it}) &= \textbf{x}_{it}^\top \bs{\beta}_2 + \textbf{b}^\top_i \bs{\delta},
\end{align*}
where $\textbf{x}_{it}$ is the $p \times 1$ vector of covariates for location $i$ and time $t$, $\bs{\beta}_1$ and $\bs{\beta}_2$ are regression coefficients, $\textbf{b}_i$ is the $q \times 1$ vector consisting of the $i$th row of $\textbf{B}$, and $\bs{\gamma}$ and $\bs{\delta}$ are basis coefficients. The priors for the basis coefficients are given by
\begin{align*}
    \bs{\gamma} = (\gamma_1,\dots,\gamma_q)^\top \mid \kappa &\sim \textup{Normal}_{q} \left( \textbf{0}, \textbf{Q}_{\text{B}}^{-1} / \kappa \right)  \\
    \bs{\delta} = (\delta_1,\dots,\delta_q)^\top \mid \tau &\sim \textup{Normal}_{q} \left( \textbf{0}, \textbf{Q}_{\text{B}}^{-1} / \tau \right), 
\end{align*}
where $\textbf{Q}_{\text{B}} = \textbf{B}^{\top} \textbf{Q} \textbf{B}$. 


\section{A fast Monte Carlo algorithm}
\label{sec:algorithm}

Here we propose a fast Monte Carlo algorithm to provide fast and reliable inference for the spatial COMP$_{\mu}$ models. We approximate $\lambda(\mu, \nu)$ using a fast spline interpolation \citep{Akima1996} and use the approximation for each accept-reject ratio of the proposed algorithm. 
The indicator variables $I_{\gamma_{j}}$ and $I_{\delta_{j}}$ for spatial random effects $\gamma_j$ and $\delta_j$ are introduced for basis-vector selection. We employ an exchange algorithm to sidestep evaluating the intractable normalizing function. We begin with an outline of our algorithm for the spatial ZICOMP$_{\mu}$ regression model. This can easily apply to the spatial COMP$_{\mu}$ regression model. 

\begin{enumerate}[Step 1.]
    \item Use an exchange algorithm for $w_{it}$ for $i = 1, \dots, n$ and $t=1,\dots,T$.
    \item Use MH updates for $\bs{\beta}_1$, $\bs{\gamma}$, and $I_{\gamma_{j}}$ for $j = 1,\dots,q$.
    \item Use an exchange algorithm for $\bs{\beta}_2$, $\nu$, $\bs{\delta}$, $I_{\delta_{j}}$ for $j = 1, \dots, q$.
    \item Do Gibbs updates for $\kappa$ and $\tau$.
\end{enumerate}

The proposals for auxiliary variables in Step 1 and Step 3 are given by \eqref{eq:auxgenw} in Section~\ref{subsec:exchange} and the spatial ZICOMP$_{\mu}$ regression model, respectively. We can generate the auxiliary variables in parallel in Steps 1 and 3. We provide details in the following sections.

\subsection{Spline approximation to rate function}
\label{subsec:rate}

To expedite fitting the models, we replace $\lambda(\mu, \nu)$ with a fast spline approximation. Consider $\Psi = [\log(0.01), \log(\mu_{\max})] \times [0.01, \nu_{\max}]$ where $\mu_{\max}$ and $\nu_{\max}$ represent the upper bounds of mean and dispersion, respectively. We set the $\mu_{\max}$ to $2 y_{\max}$ where $y_{\max}$ is the maximum observed response. The $\nu_{\max}$ is set to 5, which is more than twice $\nu$ across the simulations and applications. 
We generate a number of particles over $\Psi$ via a quasi-random number generator \citep{Faure2009} using R package \texttt{qrng}. We observed that the quasi-random sampling can provide a smaller root mean squared error (RMSE) between $\log \lambda$ and $\widehat{\log \lambda}$ than other sampling methods given the same number of particles. Consider $d$ particles $(\log \mu^{(1)}, \nu^{(1)})$, $\dots$, $(\log \mu^{(d)}, \nu^{(d)})$ in $\Psi$.
For $j = 1,\dots,d$, we find $\lambda_{\text{NR}} (\mu^{(j)}, \nu^{(j)})$ by truncating the infinite sum in (1) at the level of $\max\{2000, 10\mu^{(j)}\}$ and solving 
\begin{align*}
    \sum_{y=0}^{\max\{2000, 10\mu^{(j)}\}} (y - \mu^{(j)}) \frac{\lambda^{y}}{(y!)^{\nu^{(j)}}} = 0
\end{align*}
with respect to $\lambda$ via the NR algorithm. We chose the truncation level conservatively to provide as accurate approximations as possible. We use a very well implemented NR algorithm provided in R package \texttt{mpcmp} that provides a single solution close to the truth in every simulated example we considered. Given the particles and associated NR estimates, we approximate $\log \lambda(\mu, \nu)$ via \citet{Akima1996}'s algorithm. This algorithm divides $\Psi$ into a number of triangular cells the vertices of which are located at the NR estimates. The function values in each triangle is interpolated by a bivariate fifth-degree polynomial in $\log\mu$ and $\nu$, which allows for approximation of $\log \lambda(\mu, \nu)$ at other values of $\mu$ and $\nu$. We provide more detail of the algorithm in Supplemental Section A.1. \citet{Akima1996}'s algorithm can be carried out easily using R package $\texttt{akima}$ \citep{Gebhardt2022akima}.

We implemented simulation experiments for the proposed approximation approach, which can be found in Supplemental Sections A.2 and A.3. We found that choosing $d = \lfloor \log(\mu_{\max}) \times \nu_{\max} \times 15 \rfloor$ can provide an RMSE small enough to provide reliable approximations. The preliminary step (i.e., generating particles and finding associated NR estimates) for spline approximation only takes a couple of minutes using a single core across simulations and applications.

Our approach can dramatically reduce computing time in fitting models 
because the spline approximation is extremely fast per iteration of MCMC. For a new point $(\mu^{\ast}, \nu^{\ast})$ the triangle in which the new point lies is determined. The approximation is given by plugging in the new point into the fifth-degree polynomial. This provides a significant gain in computational efficiency over the NR method.

\subsection{RJMCMC for basis-vector selection}
\label{subsec:rjmcmc}

We propose using RJMCMC to allow for automatic selection of suitable basis vectors for the spatial effects. We introduce latent variables indicating whether basis vectors are included in the model. 
Now the models for $\pi_{it}$ and $\mu_{it}$ are given by
\begin{align*}
    \text{logit}(\pi_{it}) &= \textbf{x}_{it}^\top \bs{\beta}_1 + \textbf{b}^\top_i ( \bs{I}_{\bs{\gamma}} \bs{\gamma} )\\
    \log(\mu_{it}) &= \textbf{x}_{it}^\top \bs{\beta}_2 + \textbf{b}^\top_i ( \bs{I}_{\bs{\delta}} \bs{\delta} ),
\end{align*}
where $\bs{I}_{\bs{\gamma}}$ = diag$[(I_{\gamma_{1}}, \dots, I_{\gamma_{q}})^{\top}]$ and $\bs{I}_{\bs{\delta}}$ = diag$[(I_{\delta_{1}}, \dots, I_{\delta_{q}})^{\top}]$. The latent variables $I_{\gamma_{j}}$ and $I_{\delta_{j}}$ are 1 if the $j$th basis vector is suitable, and they are 0 otherwise. It can be computationally demanding to update $I_{\gamma_j}$ and $I_{\delta_j}$ for $j = 1, \dots, q$ at every iteration for big data. To speed computation, we could randomly select a subset of the basis vectors during each iteration and update only those indicator variables. Alternatively, we could update all variables at every $k$th iteration. To our knowledge, no existing theory suggests that the latter approach is asymptotically exact. But we found that this method produces faster convergence than the former method and correctly chooses important basis vectors. And so we use the latter method with $k = 200$ in the sequel.

\subsection{The exchange algorithm}
\label{subsec:exchange}

Let $\bs{y} = (y_{11}, \dots, y_{nT})^\top$, and let $\bs{\theta} = (\bs{w}^\top, \bs{\beta}_1^\top, \bs{\beta}_2^\top, \nu, \bs{\gamma}^\top, \bs{\delta}^\top, I_{\gamma_1}, \dots, I_{\gamma_q}, I_{\delta_1}, \dots, I_{\delta_q}, \kappa, \tau)^\top$ be the collection of model parameters. Then the joint posterior distribution of $\bs{\theta}$ is given by
\begin{align*}
    \pi(\bs{\theta} \mid \bs{y}) & \propto p(\bs{\theta}) L(\bs{\theta} \mid \bs{y}) = p(\bs{\theta}) \prod_{i=1}^n \prod_{t=1}^T  (1-\pi_{it})^{1-w_{it}} \left\{ \frac{\pi_{it}}{c[\lambda(\mu_{it}, \nu), \nu]} \frac{\lambda(\mu_{it}, \nu)^{y_{it}}}{(y_{it} !)^{\nu}} \right\}^{w_{it}}, 
\end{align*} 
where $p(\bs{\theta})$ denotes a prior and $L(\bs{\theta} \mid \bs{y})$ represents the likelihood of our model. Let $\bs{\theta}_{k}$ denote a subset of the parameters and $\bs{\theta}_{-k}$ denote the rest. The full conditional posterior of $\bs{\theta}_{k}$ is given by $\pi(\bs{\theta}_{k} \mid \bs{\theta}_{-k}, \bs{y}) \propto p(\bs{\theta}_{k}) L(\bs{\theta}_{k} \mid \bs{\theta}_{-k}, \bs{y})$,
where $p(\bs{\theta}_k)$ is a prior and $L(\bs{\theta}_{k} \mid \bs{\theta}_{-k}, \bs{y})$ is obtained by removing all terms not involving $\bs{\theta}_k$ from $L(\bs{\theta} \mid \bs{y})$.
Now consider $\pi(\bs{\theta}_{k} \mid \bs{\theta}_{-k}, \bs{y})$. The MH algorithm proposes $\bs{\theta}^{\prime}_{k}$ from $q(\cdot \mid \bs{\theta}_{k})$ and accepts $\bs{\theta}^{\prime}_{k}$ with probability
\begin{align*}
    \alpha (\bs{\theta}^{\prime}_{k} \mid \bs{\theta}_{k}) = \min \left\{ 1, \frac{ p(\bs{\theta}^{\prime}_{k}) L(\bs{\theta}^{\prime}_{k} \mid \bs{\theta}_{-k}, \bs{y}) q(\bs{\theta}_{k} \mid \bs{\theta}^{\prime}_{k}) }{ p(\bs{\theta}_{k}) L(\bs{\theta}_{k} \mid \bs{\theta}_{-k}, \bs{y})q(\bs{\theta}^{\prime}_{k} \mid \bs{\theta}_{k}) } \right\}, 
\end{align*}
at each step of the algorithm. 
We can use MH updates for $\bs{\beta}_1$, $\bs{\gamma}$, and $\bs{I}_{\bs{\gamma}}$ and Gibbs updates for $\kappa$ and $\tau$.

Standard MCMC cannot be used for the other parameters, however. Consider parameters besides $\bs{\beta}_1$, $\bs{\gamma}$, $\bs{I}_{\bs{\gamma}}$, $\kappa$, and $\tau$. Then the full conditional of $\bs{\theta}_k$ is given by
\begin{align*}
 \pi(\bs{\theta}_{k} \mid \bs{\theta}_{-k}, \bs{y}) &\propto p(\bs{\theta}_{k}) \prod_{i=1}^n \prod_{t=1}^T (1-\pi_{it})^{1-w_{it}} \left\{ \frac{\pi_{it}}{c[\lambda(\mu_{it}, \nu), \nu]} \frac{\lambda(\mu_{it}, \nu)^{y_{it}}}{(y_{it} !)^{\nu}} \right\}^{w_{it}}.
\end{align*}
Let $h(\bs{y} \mid \bs{\theta})$ = $\prod_{i=1}^n$ $\prod_{t=1}^T$ $\{ \lambda(\mu_{it}, \nu)^{y_{it}} / (y_{it}!)^{\nu} \}^{w_{it}}$ be the unnormalized likelihood and  $r(\bs{\theta})$ = $\prod_{i=1}^n$ $\prod_{t=1}^T$ $(1-\pi_{it})^{1-w_{it}}$ $\left\{ \pi_{it} / c[\lambda(\mu_{it}, \nu), \nu] \right\}^{w_{it}}$ be its normalizing function, which is intractable. 
The MH acceptance probability 
becomes
\begin{align*}
    \alpha (\bs{\theta}^{\prime}_{k} \mid \bs{\theta}_{k}) = \min \left\{ 1, \frac{ p(\bs{\theta}^{\prime}_{k}) h( \bs{y} \mid \bs{\theta}^{\prime}) r(\bs{\theta}) q(\bs{\theta}_{k} \mid \bs{\theta}^{\prime}_{k}) }{ p(\bs{\theta}_{k}) h(\bs{y} \mid \bs{\theta}) r(\bs{\theta}^{\prime}) q(\bs{\theta}^{\prime}_{k} \mid \bs{\theta}_{k}) } \right\}, 
\end{align*}
where $\bs{\theta}^{\prime} = (\bs{\theta}^{\prime\top}_{k}, \bs{\theta}_{-k}^\top)^\top$. In the acceptance probability, intractable $r(\bs{\theta})$ does not cancel out. Thus standard MCMC techniques cannot be applied. 

\citet{murray2006} introduced an auxiliary variable $\bs{z}$ that follows $h(\bs{z} \mid \bs{\theta}^{\prime}) / r( \bs{\theta}^{\prime})$ so that the intractable terms cancel out in the MH acceptance probability. Consider the full conditional $\pi(\bs{\theta}_{k} \mid \bs{\theta}_{-k}, \bs{y})$. 
The exchange algorithm proceeds as follows: given $\bs{\theta}_{k}$,
\begin{enumerate}[1.]
    \item propose $\bs{\theta}^{\prime}_{k} \sim q(\cdot \mid \bs{\theta}_{k})$,
    \item generate an auxiliary variable exactly from the probability model at $\bs{\theta}^{\prime}$: 
    $\bs{z} \sim \frac{h(\cdot \mid \bs{\theta}^{\prime})}{r( \bs{\theta}^{\prime})}$, and
    \item accept $\bs{\theta}^{\prime}_{k}$ with probability
    \begin{align*}
        \alpha = \min \left\{ 1, \frac{ p(\bs{\theta}^{\prime}_{k}) h(\bs{y} \mid \bs{\theta}^{\prime}) \bcancel{r(\bs{\theta})} h(\bs{z} \mid \bs{\theta}) \bcancel{r(\bs{\theta}^{\prime})} q(\bs{\theta}_{k} \mid \bs{\theta}^{\prime}_{k}) }{ p(\bs{\theta}_{k}) h(\bs{y} \mid \bs{\theta}) \bcancel{r(\bs{\theta}^{\prime})} h(\textbf{z} \mid \bs{\theta}^{\prime}) \bcancel{r(\bs{\theta})} q(\bs{\theta}^{\prime}_{k} \mid \bs{\theta}_{k}) } \right\}.
    \end{align*}
\end{enumerate}
We see that $r(\bs{\theta})$ cancels in the acceptance probability. We note that the exchange algorithm provides asymptotically exact estimates of model parameters. We use the exchange algorithm for $\bs{\beta}_2$, $\nu$, $\bs{\delta}$, and $\bs{I}_{\bs{\delta}}$. A fast rejection sampling scheme for COMP distributions \citep{Chanialidis2018,Benson2021} can be used for generating the auxiliary variable in Step 2. When there is severe overdispersion, however, this method may produce poor mixing chains for the chance indicator variables $\bs{w}$. In the following section we introduce a new proposal distribution for $\bs{w}$ to address this problem.

\subsubsection{New proposal distribution for chance indicator variables}

If $y_{it}$ $>$ 0, then $w_{it}$ = 1 with probability 1, by definition. If $y_{it}$ = 0, then we observe either a structural zero (implying $w_{it}$ = 0) or a chance zero (implying $w_{it}$ = 1). Conditional on $y_{it}$ = 0, the full conditional of $\bs{\theta}_k$ = $w_{it}$ is given by
\begin{align*}
    \pi(w_{it} \mid \bs{\theta}_{-k}, y_{it} = 0) &\propto p(w_{it})(1-\pi_{it})^{1-w_{it}}  \left\{ \frac{\pi_{it}}{c[\lambda(\mu_{it}, \nu), \nu]} \right\} ^{w_{it}}, 
\end{align*}
where $p(w_{it})$ is a prior. Given $w_{it} = 1$, the full conditional is proportional to intractable $c[\lambda(\mu_{it}, \nu), \nu]$. We propose $w^{\prime}_{it}$ from the swapping distribution $q(w^{\prime}_{it} \mid w_{it}) = \delta_{w^{\prime}_{it}}(1-w_{it})$, where $\delta$ denotes the Dirac delta function. Suppose we generate an auxiliary variable according to
\begin{align*}
    z_{it} &\left\{ \begin{array}{ll}
        = 0 & \mbox{if } w^{\prime}_{it} = 0\\
        \sim \textup{COMP}_{\mu}(\mu_{it}, \nu) & \mbox{if } w^{\prime}_{it} = 1.
    \end{array} \right.  
\end{align*}
Suppose $w_{it} = 0$. Then the algorithm proposes $w^{\prime}_{it}$ = $1$, generates an auxiliary variable $z_{it}$ $\sim$ $\textup{COMP}_{\mu}(\mu_{it}, \nu)$, and accepts $w^{\prime}_{it}$ = $1$ with probability
\begin{align*}
    \alpha(w^{\prime}_{it} = 1 \mid w_{it} = 0) &= \min \left\{ 1, \frac{ p(w^{\prime}_{it}) \frac{\pi_{it}}{\bcancel{c[\lambda(\mu_{it}, \nu), \nu]}} \delta(z_{it}) }{ p(w_{it}) (1-\pi_{it}) \frac{1}{\bcancel{c[\lambda(\mu_{it}, \nu), \nu]}} \frac{\lambda(\mu_{it}, \nu)^{z_{it}}}{(z_{it} !)^{\nu}}  } \right\}, \\
    &= \min \left\{ 1, \frac{ p(w^{\prime}_{it})\pi_{it} \delta(z_{it}) }{ p(w_{it}) (1-\pi_{it}) \frac{\lambda(\mu_{it}, \nu)^{z_{it}}}{(z_{it} !)^{\nu}} } \right\}.  
\end{align*}
The acceptance probability $\alpha(w^{\prime}_{it} = 1 \mid w_{it} = 0)$ = $0$ whenever $z_{it} > 0$. In practice, when there is severe overdispersion, the probability of accepting $w^{\prime}_{it}$ = 1 becomes very small, leading to an impractical algorithm. We can address this problem by introducing the following mixture distribution for the auxiliary variable:
\begin{align}
    z_{it} &\sim \left\{ \begin{array}{ll}
        \textup{NB}(\mu_{it}, \nu) & \mbox{if } w^{\prime}_{it} = 0\\
        \textup{COMP}_{\mu}(\mu_{it}, \nu) & \mbox{if } w^{\prime}_{it} = 1,
    \end{array} \right. \label{eq:auxgenw}
\end{align}
where $\textup{NB}(a, b)$ denotes the negative binomial distribution with mean $a$ and dispersion $b$. Let $\bs{\theta}^{\prime}$ = $(w^{\prime}_{it}, \bs{\theta}_{-k}^\top)^\top$. To simplify the description of our acceptance ratio, let
\begin{align*}
    g(z_{it} \mid \bs{\theta}^{\prime}) &= \left\{ \frac{\Gamma(z_{it} + \nu)}{\Gamma(z_{it} + 1) \Gamma(\nu)} \left( \frac{\nu}{\mu_{it} + \nu} \right)^{\nu} \left( \frac{\mu_{it}}{\mu_{it} + \nu} \right)^{z_{it}} \right\}^{1-w^{\prime}_{it}} \left\{ \frac{\lambda(\mu_{it}, \nu)^{z_{it}}}{(z_{it} !)^{\nu}} \right\}^{w^{\prime}_{it}}. 
\end{align*}
An exchange algorithm with the proposal \eqref{eq:auxgenw} accepts $w^{\prime}_{it}$ with probability
\begin{align*}
    \alpha(w^{\prime}_{it} \mid w_{it}) = \min \left\{ 1, \frac{ p(w^{\prime}_{it}) (1-\pi_{it})^{1-w^{\prime}_{it}} \pi^{w^{\prime}_{it}}_{it}  g(z_{it} \mid \bs{\theta} ) }{ p(w_{it}) (1-\pi_{it})^{1-w_{it}} \pi^{w_{it}}_{it} g(z_{it} \mid \bs{\theta}^{\prime} ) } \right\}. 
\end{align*}
We found that this algorithm performed well in our simulation experiments. 

\subsection{Theoretical justification}
\label{subsec:theory}

We examine the approximation error of the proposed Monte Carlo approach in terms of total variation distance. Let $\textbf{P}$ denote the transition kernel of a Markov chain simulated by the exchange algorithm with $\pi(\bth \mid \by)$ as its stationary distribution. By replacing $\lambda(\mu_{it}, \nu)$ with $\hat{\lambda}_{\text{AK}}(\mu_{it}, \nu)$ in the acceptance probability for \textbf{P}, we obtain the approximated transition kernel $\hat{\textbf{P}}_{\text{AK}}$. The theorem below quantifies the total variation distance between $\pi(\bth \mid \by)$ and $\hat{\textbf{P}}_{\text{AK}}$.
\begin{theorem}
    Consider a Markov transition kernel $\hat{\textbf{P}}_{\text{AK}}$ constructed by plugging in $\hat{\lambda}_{\text{AK}}(\mu_{it}, \nu)$ into the acceptance probability for \textbf{P}. Let $n_{\text{NR}}$ denote the number of NR iterations. Under generally satisfied assumptions (see Supplemental Section B), we have $\Vert \pi(\cdot) - \delta_{\bth_0} \hat{\textbf{P}}^n_{\text{AK}} \Vert \leq C \rho^n + \epsilon(n_{\text{NR}}) + \epsilon(d)$ almost surely for bounded constant $C$ and $0 < \rho < 1$. \label{th}
\end{theorem}

The proof of Theorem~\ref{th} is provided in Supplemental Section B. When the theorem holds, the Markov chain samples from our algorithm will get closer to $\pi(\bth \mid \by)$ as the number $n_{\text{NR}}$ of NR iterations and the number $d$ of particles increase ($\epsilon(n_{\text{NR}})$ and $\epsilon(d)$ goes to 0 as $n_{\text{NR}}$ and $d$ increases, respectively). We note that it may be interesting to understand how it scales with $n_{\text{NR}}$ and $d$, but this is highly problem-specific and present difficulties.

\section{Simulation study}
\label{sec:sim}

Here we apply our proposed methodology to data simulated from spatial COMP$_{\mu}$ and spatial ZICOMP$_{\mu}$ regression models. An aim of the simulation experiments is to assess how our model and computational approach perform in the context of under- and overdispersed count data with spatial dependence. The underlying graph for the data is the 30 $\times$ 30 lattice. Our design matrix is $\textbf{X}_t$ = $[\textbf{x}_1 \; \textbf{x}_2]$ for $t =1,\dots,T$, where $\textbf{x}_1$ = $(x_{1,1}, \dots, x_{1,900})^{\top}$ and $\textbf{x}_2$ = $(x_{2,1}, \dots, x_{2,900})^{\top}$ are the x- and y-coordinates of the vertices. We restrict the coordinates to the unit square. We use the first 25 eigenvectors of $M$ to simulate data for our study, i.e., dim$(\bs{\gamma})$ = dim$(\bs{\delta})$ = 25 and \textbf{B} is 900 $\times$ 25.  

We assume independent Normal$(\textbf{0}, 100\,\textbf{I})$ priors for the fixed effects $\bs{\beta}$, $\bs{\beta}_1$, and $\bs{\beta}_2$, and a Normal$(0, 100)$ prior for $\log(\nu)$. We assign gamma priors with shape = 0.001 and rate = 1,000 to $\kappa$ and $\tau$. We assume that $w_{it}$ $\sim$ Bernoulli$(0.5)$ for $i=1,\dots,n$ and $t=1,\dots,T$. We assign independent Bernoulli$(0.1)$ priors to $I_{\gamma_j}$ and $I_{\delta_j}$ for $j = 1,\dots,q$. This prior is appealing since it corresponds to the prior belief that the fixed effects are sufficient to explain the data, and this prior can prevent our method from producing artifactual spatial structure in the posterior.

We generate posterior sample paths of length one million in all cases to ensure that the Monte Carlo standard errors, which we compute using the batch means method \citep{Jones2006,Flegal2008}, are sufficiently small. 
We use normal proposals for $\bs{\beta}$, $\bs{\beta}_1$,  $\bs{\beta}_2$, and $\log(\nu)$ and adapt proposal covariance matrices using the Log-Adaptive Proposal algorithm \citep{Shaby2011}. We use swapping proposals for $w_{it}$, $I_{\gamma_j}$, and $I_{\delta_j}$.

\subsection{Spatial count data}
\label{subsec:simCOMP}

\begin{figure}[t]
\centering
\includegraphics[width = 0.75\textwidth]{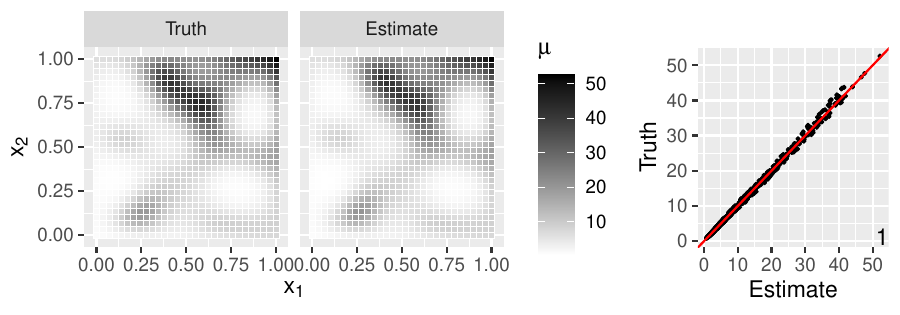}
\caption{Maps of true and estimated means $\mu_i$, and a scatter plot with the estimated correlation coefficient in the bottom right for a simulated underdispersed count dataset.\label{fig:scompUnderMap}}
\end{figure}

We create data by first setting $\tau$ = 0.2 and simulating $\bs{\delta}$ $\sim$ Normal$_{25}(\textbf{0}, \textbf{Q}_{B}^{-1}/\tau)$. For $i = 1,\dots,900$, we generate $y_i \sim \text{COMP}_{\mu}(\mu_i, \nu)$, where $\log(\mu_i) = x_{1,i} \beta_1 + x_{2,i} \beta_2 + \textbf{b}^\top_i \bs{\delta}$, $\beta_1 = 2$, and $\beta_2$ = 2. 
We generate underdispersed data by setting $\nu$ = 1.7 and overdispersed data by setting $\nu$ = 0.7. We fit our COMP$_{\mu}$ regression model with the first 100 eigenvectors of $M$ to the simulated data. We use more basis vectors than the truth to assess how our basis-vector selection approach performs. 

We see that the estimated posterior medians of parameters are close to the true values. All of their 95\% highest posterior density (HPD) intervals cover the true values (Supplemental Table 1). Figure~\ref{fig:scompUnderMap} shows that the estimated spatial pattern closely mirrors the true spatial pattern for the underdispersed data. This shows that our approach recovers well the underlying spatial pattern in the data. Similar results are observed for the overdispersed setting (Supplemental Figure 3).
We also examined posterior probabilities for basis-vector selection. We observed that all significant basis vectors (which have large true basis coefficients) have large posterior probabilities of inclusion (Supplemental Figures 4 and 5).
This shows that our method performs reliably in selecting important basis vectors. 

We examine parameter recovery of our model to several classical count models such as Poisson, Poisson-lognormal (PL), and negative binomial (NB) models. We simulate 100 datasets from each of the following five settings: (i) severe underdispersion: $y_i \sim \text{COMP}_{\mu} (\mu_i, \nu = 3.5)$, (ii) mild underdispersion: $y_i \sim \text{COMP}_{\mu} (\mu_i, \nu = 1.7)$, (iii) equidispersion: $y_i \sim \text{Poisson}(\mu_i)$, (iv) mild overdispersion: $y_i \sim \text{NB} (\mu_i, \theta = 2)$ with mean $\mu_i$ and variance $\mu_i + \mu_i^2 / \theta$, and (v) severe overdispersion: $y_i \sim \text{NB} (\mu_i, \theta = 0.2)$ where $\log(\mu_i) = x_{1,i} \beta_1 + x_{2,i} \beta_2 + \textbf{b}^\top_i \bs{\delta}$, $\beta_1 = 2$, and $\beta_2$ = 2. To each simulated dataset, we fit the spatial $\text{COMP}_{\mu}$, Poisson, PL, and NB regression models.

\setlength{\tabcolsep}{5pt}
\begin{table}[t]
\caption{Coverage rates (\%) of 95\% HPD intervals for the regression coefficients and dispersion parameter for the spatial COMP$_{\mu}$, Poisson, Poisson-lognormal (PL), and negative binomial (NB) regression models under severe under-, mild under-, equi-, mild over-, and sever overdispersion.\label{tab:coverage}}
\begin{center}
\begin{tabular}{l rr c rr c rrr rrr c }
    \hline
    \multirow{3}{*}{Generative model} & \multicolumn{13}{c}{Fitting model}\\
    \cline{2-14}
    & \multicolumn{2}{c}{Poisson} && \multicolumn{2}{c}{PL} && \multicolumn{3}{c}{NB} && \multicolumn{3}{c}{COMP$_{\mu}$}\\
    \cline{2-3} \cline{5-6} \cline{8-10} \cline{12-14}
    & $\beta_1$ & $\beta_2$ && $\beta_1$ & $\beta_2$ && $\beta_1$ & $\beta_2$ & $\theta$ && $\beta_1$ & $\beta_2$ & $\nu$\\
    \hline
    $\text{COMP}_{\mu} (\mu_i, \nu = 3.5)$---severe under & 75 & 86 &&  8 & 28 && 73 & 86 & - && 87 & 92 & 97 \\
    $\text{COMP}_{\mu} (\mu_i, \nu = 1.7)$---mild under & 86 & 85 && 32 & 31 && 81 & 86 & - && 95 & 94 & 98 \\ 
    $\text{Poisson}(\mu_i)$---equi & 76 & 79 && 54 & 51 && 76 & 78 & - && 80 & 83 & 98 \\ 
    $\text{NB} (\mu_i, \theta = 0.2)$---mild over & 83 & 84 && 79 & 78 && 88 & 93 & 97 && 94 & 95 & - \\ 
    $\text{NB} (\mu_i, \theta = 2)$---severe over & 59 & 48 && 91 &  4 && 68 & 63 & 87 && 62 & 70 & -\\ 
    \hline
\end{tabular}
\end{center}
\end{table}
Table~\ref{tab:coverage} shows the coverage rates of 95\% HPD intervals for regression coefficients and dispersion parameter. When data exhibit severe or mild underdispersion, the COMP$_{\mu}$ model best recovers the truth. The Poisson and NB models provide similar results. The PL model performs worst. Under equidispersion, the Poisson, NB, and COMP$_{\mu}$ models provide similar coverage rates, which are somewhat smaller than the nominal level. The PL model yields the smallest coverage rates. Under mild overdispersion, the NB and COMP models have coverage rates close to the nominal level. The Poisson and PL models provide relatively small coverage rates. When counts are severely overdispersed, the NB and COMP models have the best coverage rates but they are considerably smaller than the nominal level. None of the datasets were generated from the PL model. It is not unexpected that the PL model yields biased estimates for the coefficients. We note that the COMP$_{\mu}$ model performs well in parameter recovery, regardless of how the data were generated, except for the severe overdispersed case, where none of the models performed well.

We also compare the models via the widely applicable Watanabe--Akaike information criterion \cite[WAIC;][]{Watanabe2010} for each simulation setting (Supplemental Figure 8). For data with mild or severe underdispersion, the COMP$_{\mu}$ model provides better WAIC values compared to the other models. Under equidispersion, the WAIC values are similar across the four different models. When data exhibit mild or severe overdispersion, the WAIC indicates that the PL model is the best and the Poisson model is the worst. The true model is the NB model, i.e., the overdispersed data were generated by the NB model. We observed that the PL model provides a significantly better out-of-sample predictive fit than the true (NB) model. However, the PL model has a far higher model complexity than the true and COMP$_{\mu}$ models, making the PL model more difficult to interpret. The COMP$_{\mu}$ model provides similar results to the true model. In summary, the COMP$_{\mu}$ model can deal with under- and overdispersion while maintaining the same level of parsimony and interpretability as traditional count models such as the Poisson and NB models.

\subsection{Spatial zero-inflated count data}
\label{subsec:simZICOMP}

We create data by first setting $\kappa = \tau$ = 0.2 and simulating $\bs{\gamma}$ $\sim$ Normal$_{25}(\textbf{0}, \textbf{Q}_{B}^{-1}/\kappa)$ and  $\bs{\delta}$ $\sim$ Normal$_{25}(\textbf{0}, \textbf{Q}_{B}^{-1}/\tau)$. For $i = 1,\dots,900$ and $t = 1, \dots, 6$, we generate $y_{it} \sim \text{ZICOMP}_{\mu}(\pi_{i}, \mu_i, \nu)$, where
\begin{align*}
    \text{logit}(\pi_i) &= x_{1,i} \beta_{11} + x_{2,i} \beta_{12} + \textbf{b}^\top_i \bs{\gamma}\\
    \log(\mu_i) &= x_{1,i} \beta_{21} + x_{2,i} \beta_{22} + \textbf{b}^\top_i \bs{\delta},
\end{align*}
$\beta_{11} = -2$, $\beta_{12}$ = 1, $\beta_{21} = 2$, and $\beta_{22}$ = 2. 
We generate underdispersed data by setting $\nu$ = 1.7 and overdispersed data by setting $\nu$ = 0.7. We fit our ZICOMP$_{\mu}$ regression model with the first 50 eigenvectors of $M$ to the data. We use more basis vectors than the truth to assess our basis-vector selection method in this example. 

\begin{figure}[tb]
\centering
\includegraphics[width = 0.75\textwidth]{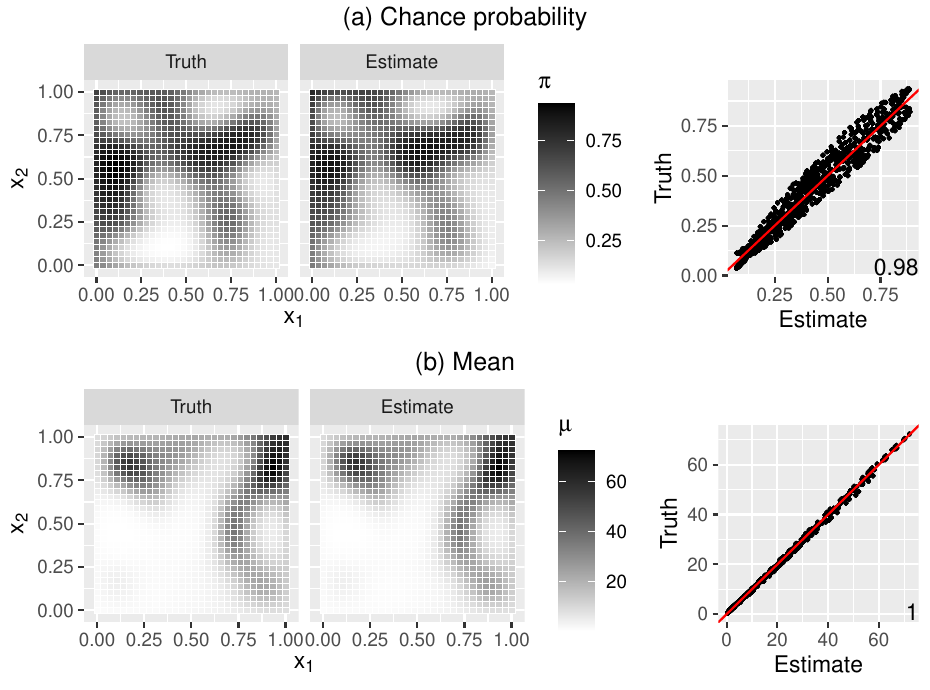}
\caption{(a) Maps of true and estimated chance probabilities $\pi_i$, and a scatter plot with the estimated correlation coefficient in the bottom right for a simulated zero-inflated underdispersed count dataset. (b) Those for means $\mu_i$. \label{fig:szicompUnderMap}}
\end{figure}

We see that the estimated posterior medians of parameters are close to the truth. All of their 95\% HPD intervals cover the truth (Supplemental Table 2). Figure~\ref{fig:szicompUnderMap} shows estimated spatial patterns closely mirror the true spatial patterns for the underdispersed data. This indicates that our approach recovers well the underlying spatial patterns in the data. Similar results are observed for the overdispersed data (Supplemental Figure 9). 
We also examined posterior probabilities for basis-vector selection. We observed that all significant basis vectors (which have large true basis coefficients) have large posterior probabilities of inclusion (Supplemental Figures 10 and 11) This implies that our methodology can successfully detect important basis vectors for both chance probability and mean.

To validate the performance more thoroughly, we apply our approach to 100 datasets. For each parameter we examine the distributions for estimated posterior medians across the 100 samples (supplementary material). The point estimates are distributed around the truth. We also estimate frequentist coverage rates based on 95\% HPD intervals. 
For the underdispersed data we observe 0.79 for $\beta_{11}$, 0.79 for $\beta_{12}$, 0.95 for $\beta_{21}$, 0.97 for $\beta_{22}$, and 0.84 for $\log(\nu)$. For the overdispersed data we observe 0.97 for $\beta_{11}$, 0.96 for $\beta_{12}$, 0.98 for $\beta_{21}$, 0.98 for $\beta_{22}$, and 0.89 for $\log(\nu)$.
This shows that $\beta_{11}$ and $\beta_{12}$ may be difficult to recover when data are underdispersed. This is so because when chance observations have means close to 0 and are underdispersed, only zero outcomes may be observed and may be considered structural zeros by the model.

\section{Application to real data}
\label{sec:real}

Here we apply our methodology to two real dataset that exhibit challenging features such as spatial dependence, zero inflation, underdispersion, or overdispersion. 

\subsection{Texas HPV-cancer data}
\label{subsec:hpv}

Data on the number of cancer cases in the 254 counties of Texas between 2006 and 2012 were obtained from the Texas Cancer Registry \citep{amber2015}. The set of covariates includes county-level demographic or socioeconomic factors (Supplemental Table 3). The dataset contains records for approximately 213 counties. 

To these data we fit our spatial COMP$_{\mu}$ regression model with the first 80 eigenvectors of $M$ (a total of 85 eigenvectors display attractive patterns). We use the total population as the offset variable. 
We generated posterior sample paths of length 1 million, which  took approximately five hours. The first 0.2 million iterations were discarded as burn-in. The estimated posterior median and 95\% HPD interval of $\nu$ are 1.2 and (1.0, 1.5), which indicates underdispersion.

We find that communities with a high risk of HPV-related cancers tend to have a high female population, contain individuals aged 15 to 19 or 30 and older, and have a high smoking rate (Supplemental Figure 14). The estimated posterior mean and 95\% HPD interval of $\tau$ are 17.7 and (1.8, 55.8), which are much smaller than the prior mean of 1,000. One basis vector was found to be a significant predictor (Supplemental Figure 15).

\subsection{US vaccine refusal data}
\label{subsec:vaccine}

Monthly reports for vaccine refusal among patients under five years of age were obtained from a database of U.S. medical claims managed by IMS Health \citep{kang2023spatial}. We examine childhood vaccine refusal from January to June in 2015. The response is the number of vaccine refusal cases in a county in a month. Approximately 84\% of the outcomes are zeros. The set of covariates includes factors representing detection mechanisms of vaccine refusal and demographic or socioeconomic factors (Supplemental Table 4). The dataset contains records for approximately 3,000 counties for each month. 

To these data we fit our spatial ZICOMP$_{\mu}$ regression model with the first 100 eigenvectors of $M$. 
We generated posterior sample paths of length 1.6 million, which took approximately 3.3 days. The first 0.8 million iterations were discarded as burn-in. 
Using the NR algorithm was computationally infeasible for these data. Our spline approximation to the rate function $\lambda(\mu_{it}, \nu)$ provided significant computational gains over the NR method. The estimated posterior median and 95\% HPD interval of $\nu$ are 0.105 (0.084, 0.126), which indicates severe overdispersion.

We find that the number of physician-patient interactions is overwhelmingly predictive of refusal detection, and healthcare insurance coverage may be predictive of refusal measurement. Among demographic or socioeconomic variables, we observe that communities with small household sizes, religions historically opposed to vaccination, high incomes, and high leniency
in the state’s vaccination are likely to have reported refusals. Given perfect detection, we observe that communities with high refusal are likely to have increased access to care, have low insurance coverage, have more physician reporting, have small household sizes, contain groups historically opposed to vaccination, more people lacking English proficiency, have high rates of private school attendance, have high incomes, lack continuity of care, have high leniency in state vaccination laws, and have high incidence of autism (Supplemental Figure 17). 

The estimated posterior mean and 95\% HPD interval of $\kappa$ and $\tau$ are 0.07 (0.01, 0.08) and 0.07 (0.05, 0.10), respectively. They are much smaller than the prior mean of 1,000. Fourteen basis vectors were found to be significant predictors of detection of refusal. Thirty-four basis vectors were found to be significant predictors for refusal cases (Supplemental Figure 18). This suggests that conducting a non-spatial analysis of these data would be inappropriate.

\section{Discussion}
\label{sec:discuss}

In this article we proposed new mean-parameterized Conway--Maxwell--Poisson regression models that can account for spatial dependence, zero inflation, underdispersion, equidispersion, or overdispersion. We also proposed several computational approaches that are efficient for carrying out Bayesian inference for our models. 
We note that the approaches of \citet{Philipson2023} and \citet{Philipson2023trunc} enable fast computation for non-spatial COMP$_{\mu}$ models; the former for data with small or moderate counts and the latter for large datasets in the form of small counts. To our knowledge, our approach is the first practical method for analyzing big spatial data with large counts using \citet{Huang2017}'s COMP$_{\mu}$, which has appealing properties over alternative models.
Our approach may be useful in many disciplines, such as ecology, agriculture, criminology, medicine, and public health studies where spatial count data are commonly encountered. Due to high-dimensional spatial effects, Markov chains can mix slowly, which inspired us to use basis representation techniques. However, we still need to generate long sample paths to ensure convergence of the chains. Employing the Hamiltonian Monte Carlo algorithm \citep{duane1987hybrid,neal2011mcmc} for faster convergence may be an interesting topic for future research. 



\section*{Acknowledgments}

This work was supported by the National Institute of General Medical Sciences of the National Institutes of Health under Award Number R01GM123007. We are grateful to Shweta Bansal for helpful conversations.

\bibliographystyle{chicago}
\bibliography{refs}

\end{document}


\def\spacingset#1{\renewcommand{\baselinestretch}%
{#1}\small\normalsize} \spacingset{1}
\newcommand{\bs}{\boldsymbol}
\newcommand{\bw}{\bs{w}}
\newcommand{\bx}{\bs{x}}
\newcommand{\by}{\bs{y}}
\newcommand{\bz}{\bs{z}}
\newcommand{\bu}{\bs{u}}
\newcommand{\bX}{\bs{X}}
\newcommand{\bY}{\bs{Y}}
\newcommand{\bZ}{\bs{Z}}
\newcommand{\bth}{\bs{\theta}}
\newcommand{\bTh}{\bs{\Theta}}
\newcommand{\bet}{\bs{\eta}}
\newcommand{\bcth}{\bs{\Theta}}

\newtheorem{definition}{Definition}
\newtheorem{assumption}{Assumption}
\newtheorem{theorem}{Theorem}
\newtheorem{proposition}{Proposition}

\def\thesection{\Alph{section}}


\if0\blind
{
  \title{\bf Supplementary Material for ``Fast Bayesian Inference for Spatial Mean-Parameterized Conway--Maxwell--Poisson Models''}
  \author[]{Bokgyeong Kang}
  \author[]{John Hughes}
  \author[]{Murali Haran}
  \affil[]{}
  \date{}
  \maketitle
} \fi

\if1\blind
{
  \bigskip
  \bigskip
  \bigskip
  \begin{center}
    {\LARGE\bf Title}
\end{center}
  \medskip
} \fi


\spacingset{1.5}

\section{Spline approximation}

Here we provide more details about our spline approximation to $\lambda(\mu, \nu)$, and simulation experiments to examine its computational cost and performance.

\subsection{Details of \citet{Akima1996}'s algorithm}

Consider $d$ particles $(\log \mu^{(1)}, \nu^{(1)})$, $\dots$, $(\log \mu^{(d)}, \nu^{(d)})$ in $\Psi = [\log(0.01), \log(\mu_{\max})] \times [0.01, \nu_{\max}]$. For $j = 1, \dots, d$ we can obtain the estimates $\lambda_{\text{NR}}(\mu^{(j)}, \nu^{(j)})$ through the NR algorithm. Let $f(\mu, \nu)$ denote $\log \lambda(\mu, \nu)$ for simplicity of notation. Given the particles and estimates, we approximate the function $f$ via \citet{Akima1996}'s algorithm as follows.
\begin{enumerate}[1.]
    \item At each particle five partial derivatives, $f_{\mu}$, $f_{\nu}$, $f_{\mu \mu}$, $f_{\nu \nu}$, and $f_{\mu \nu}$, are approximated based on Appendix A of \citet{Akima1996}.
    \item The particle area is partitioned into triangles the vertices of which are located at the NR estimates.
    \item For each triangle the function $f$ is assumed to be a bivariate fifth-degree polynomial in $\log \mu$ and $\nu$, i.e., $f(\mu, \nu) \approx \sum_{k=0}^{5} \sum_{l=0}^{5-k} a_{kl} (\log \mu)^k \nu^l$.
    Note that there are 21 coefficients to be determined for each triangle.
    \item The NR estimates of $f$ and the estimates of the five partial derivatives (i.e., $f_{\mu}$, $f_{\nu}$, $f_{\mu \mu}$, $f_{\nu \nu}$, and $f_{\mu \nu}$) are given at each vertex of the triangle. This produces 18 independent equations.
    \item We assume that the partial derivative of $f$ differentiated in the direction perpendicular to each side of the triangle is a polynomial of degree three, at most, in the variable measured in the direction of the side of the triangle. This assumption yields three independent equations (see Appendix A of \citet{Akima1978} for details). 
    \item For each triangle the coefficients $a_{kl}$ are determined by solving the system of 21 equations.
\end{enumerate}

The spline interpolation is extremely fast per iteration. For a new point $(\mu^{\ast}, \nu^{\ast})$ the triangle in which the new point lies is determined. The approximation is given by $\hat{f}(\mu^{\ast}, \nu^{\ast})$ $=$ $\sum_{k=0}^{5} \sum_{l=0}^{5-k} a_{kl} (\log \mu^{\ast})^k (\nu^{\ast})^l$. This provides a significant gain in computational efficiency over the NR method.

\subsection{Computational cost}

We conduct a simulation experiment to study the computing time of the preliminary step for function approximation. We generate a number $d$ of particles over $[\log(0.01), \log (\mu_{\max})] \times [0.01, \nu_{\max}]$ using quasi-random sampling for different combinations of $d$ and $\mu_{\max}$. The upper bound $\nu_{\max}$ of the dispersion parameter is set to 5, which is more than twice $\nu$ across the simulations and applications. We find $\lambda$ at each particle via the NR algorithm. We approximate $\log \lambda(\mu, \nu)$ through spline interpolation. Note that a single core was used here, but the algorithm is paralellizable. For each value of $\mu_{\max}$, we increase $d$ until the root mean squared error (RMSE) between $\log \lambda(\mu,\nu)$ and $\widehat{\log \lambda}(\mu,\nu)$ is small enough.
\begin{figure}[H]
    \centering
    \includegraphics[width = 0.5\textwidth]{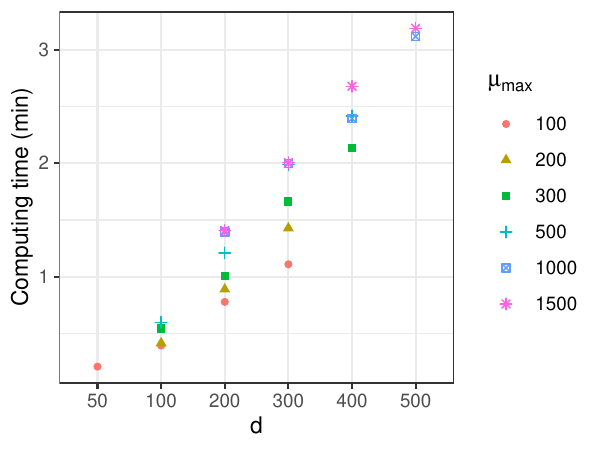}
    \caption{The computing time, averaged over 30 simulations, of the preliminary step for different combinations of $d$ and $\mu_{\max}$. Note that each $\mu_{\max}$ has a different array of $d$ values.}
    \label{sup:fig:time}
\end{figure}
Figure~\ref{sup:fig:time} shows the computing time of the preliminary step for function approximation for different combinations of $d$ and $\mu_{\max}$. Trivially, the computation time grows as $d$ increases. We see that it takes longer for larger values of $\mu_{\max}$. Finding $\lambda$ takes longer when $\mu$ values are large. We note that the preliminary step only takes approximately 3.5 minutes when $\mu_{\max}$ = 1500.

\subsection{Comparison to different approximation methods}

We implement a simulation experiment to compare the performance of the following three different approaches for approximating $\lambda(\mu, \nu)$.
\begin{enumerate} [(i)]
    \item Quasi-random sampling $+$ NR $+$ spline (our proposed method)
    \item Grid $+$ \texttt{polyroot} $+$ bilinear \citep{Philipson2023}
    \item Quasi-random sampling $+$ \texttt{polyroot} $+$ spline
\end{enumerate}
Method (i) is our proposed approach that generates a number $d$ of particles over $[\log(0.01), \log(\mu_{\max})]$ $\times$ $[0.01, \nu_{\max}]$ using quasi-random sampling, finds $\lambda$ at each particle via the NR algorithm, and approximates $\log \lambda(\mu, \nu)$ through the spline extra- and interpolation. We used a very well implemented NR algorithm provided in R package \texttt{mpcmp} that provides a single solution close to the truth in every simulated example we considered. Given the truncation level of $\max\{2000, 10 \mu\}$, the NR algorithm never broke and provided a single root. Method (ii) was introduced by \citet{Philipson2023} which creates a fine resolution grid over $[\log(0.01), \log(\mu_{\max})]$ $\times$ $[0.01, \nu_{\max}]$, finds $\lambda$ at each grid vertex via the Jenkins-Traub (JT) algorithm using \texttt{polyroot}, and approximates $\log \lambda(\mu, \nu)$ through the bilinear extra- and interpolation. Method (iii) is identical to Method (i), except that the NR algorithm is replaced with the JT algorithm.
    
The $\nu_{\max}$ set to 5, which is more than twice $\nu$ across the simulations and applications. For Methods (i) and (iii), we compute the root mean squared error (RMSE) between $\log\lambda(\mu,\nu)$ and $\widehat{\log \lambda}(\mu,\nu)$ for different combinations of $d$ and $\mu_{\max}$. For Method (ii), we create grids with the recommended step size of 0.01 and compute the RMSE with an increasing value of $\mu_{\max}$. We find that Method (iii) produces significantly large RMSE values compared to Methods (i) and (ii); the RMSE for Method (iii) ranges from 0.37 to 5.90, while the RMSEs for Methods (i) and (ii) range from 0.001 to 0.017 and 0.0002 to 0.053, respectively.
\begin{figure}[t]
    \centering
    \includegraphics[width = \textwidth]{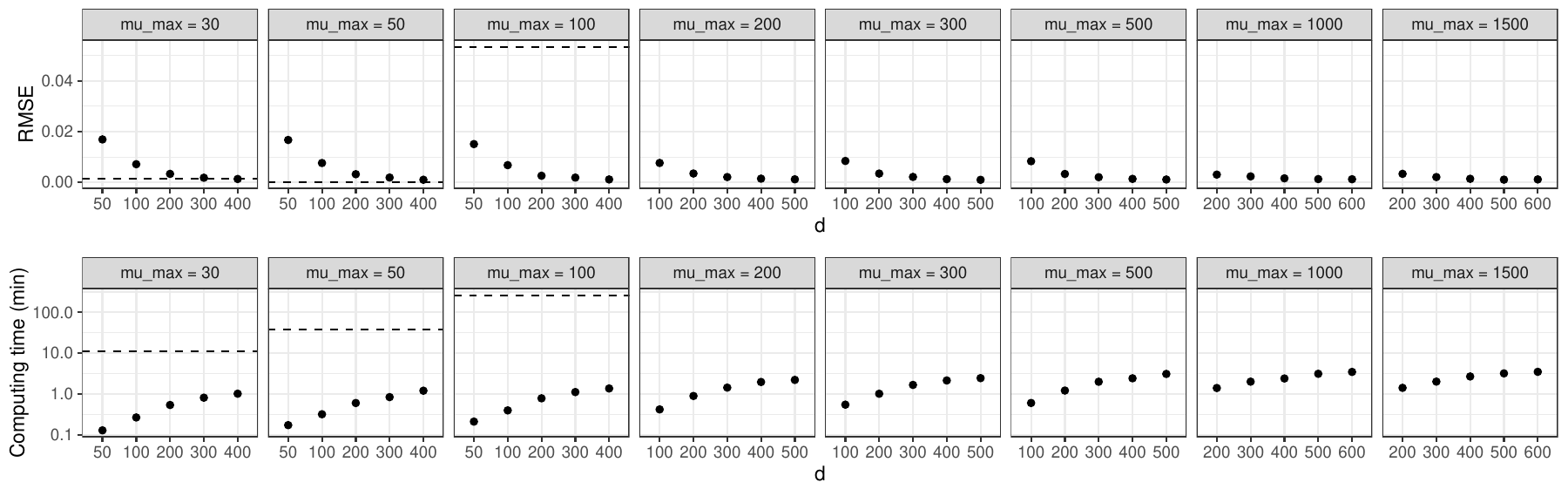}
    \caption{RMSEs and computing times for Methods (i) and (ii). The black dots represent the RMSEs and computing times, averaged over 30 replicates, for different combinations of $d$ (x-axis) and $\mu_{\max}$ (facet panel) for Method (i). The dashed horizontal lines represent the RMSEs and computing times with an increasing value of $\mu_{\max}$ (facet panel) for Method (ii). Note that Method (ii) for $\mu_{\max} \geq 200$ was excluded from the figure due to numerical instability. Each $\mu_{\max}$ has a different array of $d$ values.}
    \label{sup:fig:rmsetime}
\end{figure} 
Figure~\ref{sup:fig:rmsetime} shows the RMSEs and computing times for Methods (i) and (ii). Note that both methods can be done in parallel but here we used a single core for comparison. We see that Method (i) outperforms Method (ii) when $\mu_{\max}$ is large. The JT algorithm is numerically unstable for high-degree polynomials. The \texttt{polyroot} failed to find roots and broke for some combinations of $\mu$ and $\nu$ when $\mu_{\max} \geq 200$. We implemented Method (ii) for $\mu_{\max} \in \{30, 50, 100\}$. Method (i) can provide an RMSE small enough for any values of $\mu_{\max}$, while Method (ii) yields a relatively large RMSE for $\mu_{\max} = 100$. Method (ii) only takes at most 4 minutes for any combinations of $d$ and $\mu_{\max}$ we consider. But Method (ii) takes approximately 4.3 hours for $\mu_{\max} = 100$. In summary, our proposed method not only provides reliable approximations to $\log \lambda(\mu, \nu)$ for any values of $\mu_{\max}$ but also is computationally very efficient.

\section{Proof of Theorem 1}

Consider a posterior distribution $\pi(\bth \mid \by)$. The exchange algorithm draws samples from $\pi(\bth \mid \by)$ by simulating a Markov chain with transition kernel \textbf{P}. The acceptance ratio of the algorithm is given by
\begin{align*}
    \alpha(\bth, \bth^\prime) = \frac{p(\bth^\prime) q(\bth \mid \bth^\prime) h(\by \mid \bth^\prime) h(\bz \mid \bth)}{p(\bth) q(\bth^\prime \mid \bth) h(\by \mid \bth) h(\bz \mid \bth^\prime)}.
\end{align*}
Let $\hat{h}_{\text{NR}}$ and $\hat{h}_{\text{AK}}$ denote the NR estimate and Akima approximation respectively. By plugging in the $\hat{h}_{\text{NR}}$ into the acceptance ratio, we can construct the first-stage approximated transition kernel $\hat{\textbf{P}}_{\text{NR}}$ and obtain the corresponding acceptance ratio $\hat{\alpha}_{\text{NR}}(\bth, \bth^\prime)$. By replacing $\hat{h}_{\text{NR}}$ with $\hat{h}_{\text{AK}}$, we can construct the second-stage approximated kernel $\hat{\textbf{P}}_{\text{AK}}$ and the corresponding acceptance ratio $\hat{\alpha}_{\text{AK}}(\bth, \bth^\prime)$. 
We make the following assumptions.
\begin{assumption}
    The initial value for the NR algorithm is sufficiently close to the truth. \label{sup:ass:initial}
\end{assumption}
\begin{assumption}
    The parameter space $\bs{\Theta}$ is compact. \label{sup:ass:par}
\end{assumption}
\begin{assumption}
    There exists a constant $c_p > 1$ such that $1/c_p \leq p(\bth) \leq c_p$. \label{sup:ass:prior}
\end{assumption}
\begin{assumption}
    There exists a constant $c_q > 1$ such that $1/c_q \leq q(\bth^\prime \mid \bth) \leq c_q$. \label{sup:ass:proposal}
\end{assumption}
\begin{assumption}
    Let $\hat{h}_{\text{NR}}(\by \mid \bth)$ denote the NR estimate obtained by replacing $\lambda(\mu_{it}, \nu)$ with $\hat{\lambda}_{\text{NR}}(\mu_{it}, \nu)$ in $h(\by \mid \bth)$. There exist constants $k$ and $K$ such that $k \leq h(\by \mid \bth) \leq K$ and $k \leq \hat{h}_{\text{NR}}(\by \mid \bth) \leq K$. \label{sup:ass:h}
\end{assumption}

We use a very well implemented NR algorithm provided in R package \texttt{mpcmp} that provides a single solution close to the truth in every simulated example we considered. For many applications we may reasonably assume that the sample space is finite and the parameter space $\bs{\Theta}$ is a compact set. The Assumptions \ref{sup:ass:prior} to \ref{sup:ass:h} may be easily checked.

\subsection{Approximation error of the NR estimates}

The difference between the acceptance probabilities of $\hat{\textbf{P}}_{\text{NR}}$ and \textbf{P} is
\begin{align*}
    \vert \hat{\alpha}_{\text{NR}}(\bth, \bth^\prime) - \alpha(\bth, \bth^\prime) \vert &= \ \left\vert \frac{\hat{h}_{\text{NR}}(\by \mid \bth^\prime) \hat{h}_{\text{NR}} (\bz \mid \bth)}{\hat{h}_{\text{NR}}(\by \mid \bth) \hat{h}_{\text{NR}}(\bz \mid \bth^\prime)} - \frac{h(\by \mid \bth^\prime) h (\bz \mid \bth)}{h(\by \mid \bth) h(\bz \mid \bth^\prime)}  \right\vert \frac{p(\bth^\prime) q(\bth \mid \bth^\prime)}{p(\bth) q(\bth^\prime \mid \bth)} \\
    &\leq \epsilon(n_{\text{NR}}) \frac{p(\bth^\prime) q(\bth \mid \bth^\prime)}{p(\bth) q(\bth^\prime \mid \bth)}\\
    &\leq \epsilon(n_{\text{NR}}) c^2_p c^2_q.
\end{align*}
where $n_{\text{NR}}$ is the number of iterations of the NR algorithm. The first inequality holds from the NR method's quadratic convergence under the Assumption~\ref{sup:ass:initial} and continuous mapping theorem. The second inequality holds from the Assumptions \ref{sup:ass:prior} and \ref{sup:ass:proposal}.

Now we will show that the Markov chain with the transition kernel \textbf{P} is uniformly ergodic for a measurable set $B \subset \bTh$. Under the Assumptions \ref{sup:ass:prior} to \ref{sup:ass:h}, we can derive the lower bound of the \textbf{P} as
\begin{align*}
    \textbf{P}(\bth, B) &= \int_{B} \delta_{\bth} (d\bth^\prime) \left( 1 - \int_{B} d \bs{t} q(\bs{t} \mid \bth) \min\{1, \alpha(\bth, \bs{t})\} \right)\\
    &\quad + \int_B d\bth^\prime q(\bth^\prime \mid \bth) \min \{1, \alpha(\bth, \bth^\prime)\}\\
    &\geq \int_B d\bth^\prime q(\bth^\prime \mid \bth) \min \{1, \alpha(\bth, \bth^\prime)\} \\
    &\geq \frac{k^2}{c^2_p c^2_q K^2} \int_B d\bth^\prime q(\bth^\prime \mid \bth) \\
    &\geq \frac{k^2}{c^2_p c^3_q K^2} \int_B d\bth^\prime.
\end{align*}
By Theorems 16.0.2 and 16.2.4 of \citet{Meyn1993}, we have 
\begin{align}
    \sup_{\bth} \Vert \delta_{\bth} \textbf{P}^n - \pi(\cdot) \Vert \leq C \rho^n, \label{ineq:errorP}
\end{align}
where $C = 2$ and $\rho = 1 - k^2 / (c^2_p c^3_q K^2)$. Thus the Markov chain with \textbf{P} is uniformly ergodic for a measurable set $B \subset \bTh$.

We have shown that (i) the difference between the acceptance probabilities of $\hat{\textbf{P}}_{\text{NR}}$ and \textbf{P} is bounded and (ii) the Markov chain with the transition kernel \textbf{P} is uniformly ergodic. By Corollary 2.3 of \citet{alquier2016}, we have 
\begin{align}
    \Vert \delta_{\bth_0} \textbf{P}^n - \delta_{\bth_0} \hat{\textbf{P}}^n_{\text{NR}} \Vert \leq \epsilon(n_{\text{NR}}) c^2_p c^2_q \left( \lambda + \frac{C \rho^\lambda}{ 1 - \rho} \right) \label{ineq:errorPnr}
\end{align}
almost surely, where $\lambda = \lceil \log(1/C) / \log(\rho) \rceil$.

\subsection{Approximation error of the Akima approximation}

The difference between the acceptance probabilities of $\hat{\textbf{P}}_{\text{AK}}$ and $\hat{\textbf{P}}_{\text{NR}}$ is
\begin{align*}
    & \vert \hat{\alpha}_{\text{AK}}(\bth, \bth^\prime) - \hat{\alpha}_{\text{NR}}(\bth, \bth^\prime) \vert \\
    & = \ \left\vert \frac{\hat{h}_{\text{AK}}(\by \mid \bth^\prime) \hat{h}_{\text{AK}} (\bz \mid \bth)}{\hat{h}_{\text{AK}}(\by \mid \bth) \hat{h}_{\text{AK}}(\bz \mid \bth^\prime)} - \frac{\hat{h}_{\text{NR}}(\by \mid \bth^\prime) \hat{h}_{\text{NR}} (\bz \mid \bth)}{\hat{h}_{\text{NR}}(\by \mid \bth) \hat{h}_{\text{NR}}(\bz \mid \bth^\prime)} \right\vert \frac{p(\bth^\prime) q(\bth \mid \bth^\prime)}{p(\bth) q(\bth^\prime \mid \bth)}.
\end{align*}
Under Assumption \ref{sup:ass:par}, there exists a finite number $d$ of open balls with radius $r(d)$ that entirely cover $\bTh$. Let $\bth^{(1)}, \dots, \bth^{(d)}$ denote the centroids of the balls. Since $\hat{h}_{\text{AK}}(\by \mid \bth)$ is continuous with respect to $\bth$, there exists $\epsilon(d)$ such that 
\begin{align*}
    \vert \hat{h}_{\text{AK}}(\by \mid \bth) - \hat{h}_{\text{NR}}(\by \mid \bth) \vert < \epsilon(d)
\end{align*}
for all $r(d) > 0$ and for every $\bth \in \bTh$. By Assumptions \ref{sup:ass:prior} and \ref{sup:ass:proposal} and continuous mapping theorem, we have 
\begin{align*}
    & \vert \hat{\alpha}_{\text{AK}}(\bth, \bth^\prime) - \hat{\alpha}_{\text{NR}}(\bth, \bth^\prime) \vert \leq \epsilon(d) c^2_p c^2_q.
\end{align*}

Now we will show that the Markov chain with the transition kernel $\hat{\textbf{P}}_{\text{NR}}$ is uniformly ergodic for a measurable set $B \subset \bTh$. Under Assumptions \ref{sup:ass:prior} to \ref{sup:ass:h}, we can derive the lower bound of the $\hat{\textbf{P}}_{\text{NR}}$ as
\begin{align*}
    \hat{\textbf{P}}_{\text{NR}}(\bth, B) &= \int_{B} \delta_{\bth} (d\bth^\prime) \left( 1 - \int_{B} d \bs{t} q(\bs{t} \mid \bth) \min\{1, \hat{\alpha}_{\text{NR}}(\bth, \bs{t})\} \right)\\
    &\quad + \int_B d\bth^\prime q(\bth^\prime \mid \bth) \min \{1, \hat{\alpha}_{\text{NR}}(\bth, \bth^\prime)\}\\
    &\geq \int_B d\bth^\prime q(\bth^\prime \mid \bth) \min \{1, \hat{\alpha}_{\text{NR}}(\bth, \bth^\prime)\} \\
    &\geq \frac{k^2}{c^2_p c^2_q K^2} \int_B d\bth^\prime q(\bth^\prime \mid \bth) \\
    &\geq \frac{k^2}{c^2_p c^3_q K^2} \int_B d\bth^\prime.
\end{align*}
By Theorems 16.0.2 and 16.2.4 of \citet{Meyn1993}, we have 
\begin{align*}
    \sup_{\bth} \Vert \delta_{\bth} \hat{\textbf{P}}^n_{\text{NR}} - \hat{\pi}_{\text{NR}}(\cdot) \Vert \leq C \rho^n,
\end{align*}
where $C = 2$ and $\rho = 1 - k^2 / (c^2_p c^3_q K^2)$. Thus the Markov chain with $\hat{\textbf{P}}_{\text{NR}}$ is uniformly ergodic for a measurable set $B \subset \bTh$.

We have shown that (i) the difference between the acceptance probabilities of $\hat{\textbf{P}}_{\text{AK}}$ and $\hat{\textbf{P}}_{\text{NR}}$ is bounded and (ii) the Markov chain with the transition kernel $\hat{\textbf{P}}_{\text{NR}}$ is uniformly ergodic. By Corollary 2.3 of \citet{alquier2016}, we have 
\begin{align}
    \Vert \delta_{\bth_0} \hat{\textbf{P}}^n_{\text{NR}} - \delta_{\bth_0} \hat{\textbf{P}}^n_{\text{AK}} \Vert \leq \epsilon(d) c^2_p c^2_q \left( \lambda + \frac{C \rho^\lambda}{ 1 - \rho} \right) \label{ineq:errorPak}
\end{align}
almost surely, where $\lambda = \lceil \log(1/C) / \log(\rho) \rceil$.

\subsection{Approximation error of the proposed approach}

Ignoring the constants, by the inequalities \eqref{ineq:errorP} to \eqref{ineq:errorPak}, the approximation error for the proposed approach is given by
\begin{align*}
    \Vert \pi(\cdot) - \delta_{\bth_0} \hat{\textbf{P}}^n_{\text{AK}} \Vert &\leq \Vert \pi(\cdot) - \delta_{\bth_0} \textbf{P}^n \Vert + \Vert \delta_{\bth_0} \textbf{P}^n - \delta_{\bth_0} \hat{\textbf{P}}^n_{\text{NR}} \Vert + \Vert \delta_{\bth_0} \hat{\textbf{P}}^n_{\text{NR}} - \delta_{\bth_0} \hat{\textbf{P}}^n_{\text{AK}}\Vert \\
    &\leq C \rho^n + \epsilon(n_{\text{NR}}) + \epsilon(d)
\end{align*}
almost surely.



\section{Supplemental figures and tables for Section 4}

Here we provide supplemental figures and tables for the simulation experiments described in Section 4 of the manuscript.

\subsection{Spatial count data}

Table~\ref{sup:tab:simscomp} shows estimated posterior medians and 95\% HPD intervals for model parameters for underdispersed and overdispersed datasets. We observe that all of their 95\% HPD intervals cover the true values.

\begin{table}[H]
\caption{(a) True values, estimated posterior medians, and 95\% highest posterior density (HPD) intervals for $\beta_1$, $\beta_2$, and $\log(\nu)$ for a underdispersed dataset. (b) Those for a overdispersed dataset. The estimates are close to the true values.}
\label{sup:tab:simscomp}
\centering
\begin{tabular}{lrrr c rrr}
    \toprule
     & \multicolumn{3}{c}{(a) Under-dispersed data} && \multicolumn{3}{c}{(b) Over-dispersed data}\\
     \cmidrule{2-4} \cmidrule{6-8}
     & Truth & Median & 95\% HPD && Truth & Median & 95\% HPD \\\midrule
    $\beta_1$ & 2.00 & 2.01 & (1.82, 2.14) && 2.00 &  2.10 & (1.96, 2.27)\\
    $\beta_2$ & 2.00 & 1.98 & (1.84, 2.17) && 2.00 & 1.95 & (1.80, 2.11)\\
    $\log(\nu)$ & 0.53 & 0.59 & (0.08, 0.74) && -0.35 & -0.35 & (-0.47, -0.24) \\\bottomrule
\end{tabular}
\end{table}

\begin{figure}[H]
\centering
\includegraphics[width = 0.75\textwidth]{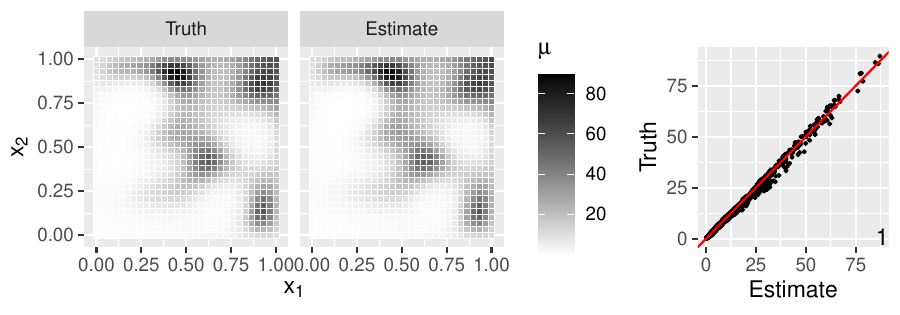}
\caption{Maps of true and estimated means $\mu_i$, and a scatter plot with the estimated correlation coefficient in the bottom right for a simulated overdispersed count dataset.\label{fig:scompOverMap}}
\end{figure}

Figure~\ref{fig:scompOverMap} shows that the estimated spatial pattern closely mirrors the true spatial pattern for the overdispersed data. This shows that our approach recovers well the underlying spatial pattern in the data. 

\begin{figure}[H]
\centering
\includegraphics[width = 0.7\textwidth]{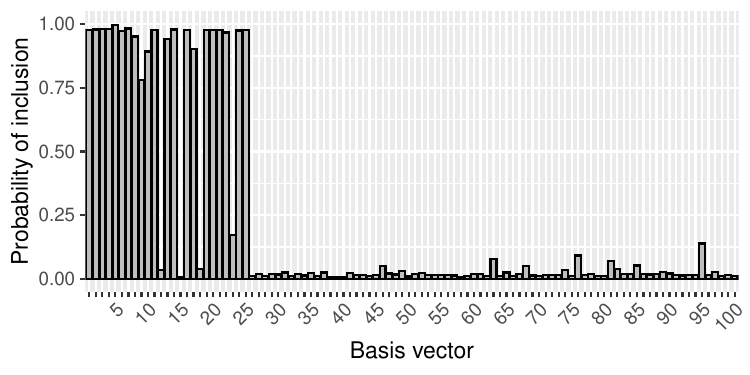}
\caption{Estimated posterior probabilities of including basis vectors for a simulated underdispersed count dataset.\label{fig:scompUnderBasis}}
\end{figure}
\begin{figure}[H]
\centering
\includegraphics[width = 0.7\textwidth]{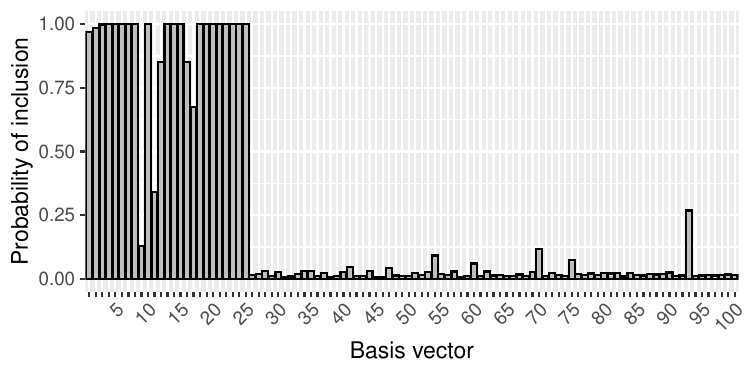}
\caption{Estimated posterior probabilities of including basis vectors for a simulated overdispersed count dataset.\label{fig:scompOverBasis}}
\end{figure}

Figures \ref{fig:scompUnderBasis} and \ref{fig:scompOverBasis} display the posterior probabilities for basis-vector selection for underdispersed dataset and overdispersed dataset, respectively. We see that all significant basis vectors (who have quite large magnitudes of true basis coefficients) have large posterior probabilities of inclusion.

\begin{figure}[H]
\centering
\includegraphics[width = 0.8\textwidth]{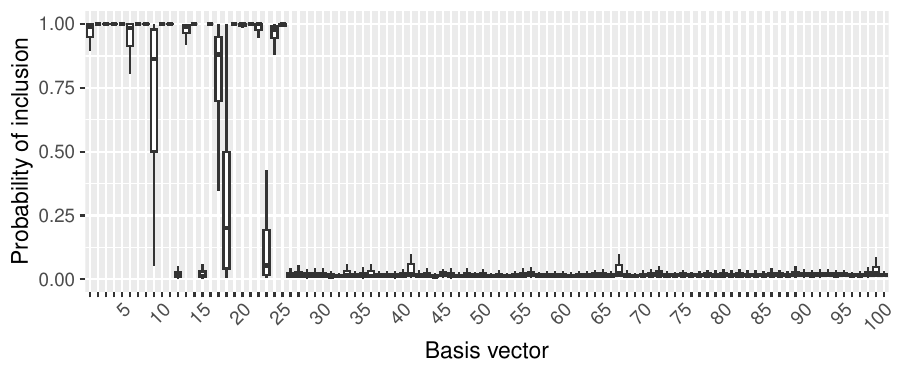}
\caption{Box plots for estimated posterior probabilities for basis-vector selection, across 100 simulations, for a simulated underdispersed count dataset.\label{fig:scompUnderBasisBox}}
\end{figure}
\begin{figure}[H]
\centering
\includegraphics[width = 0.8\textwidth]{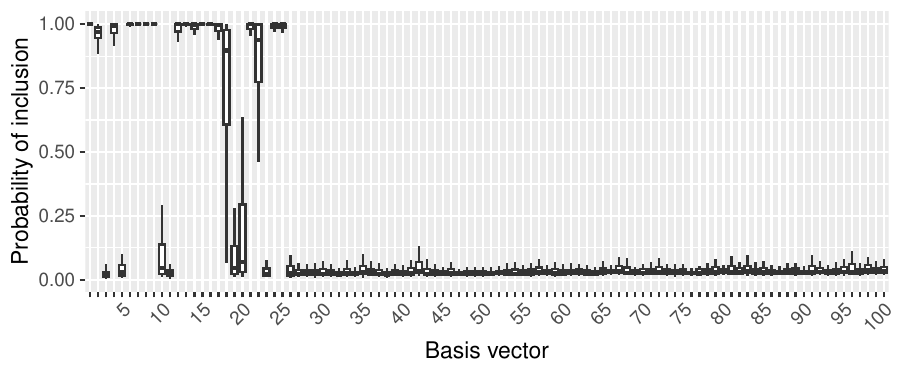}
\caption{Box plots for estimated posterior probabilities for basis-vector selection, across 100 simulations, for a simulated overdispersed count dataset.\label{fig:scompOverBasisBox}}
\end{figure}

We examine the distributions of estimated posterior means for each $I_{\delta_j}$ across the 100 samples. Figures \ref{fig:scompUnderBasisBox} and \ref{fig:scompOverBasisBox} show the results for the underdispersed and overdispersed datasets, respectively. The box plots close to 1 indicate that the associated basis vectors were selected in most cases. We find that all significant basis vectors (whose box plots are close to 1) have quite large magnitudes of true basis coefficients. The insignificant ones (whose box plots are close to 0) are found to have true basis coefficient values either close to 0 or equal to 0. This shows that our method performs reliably in selecting important basis vectors.

\begin{figure}[H]
    \centering
    \includegraphics[width = \textwidth]{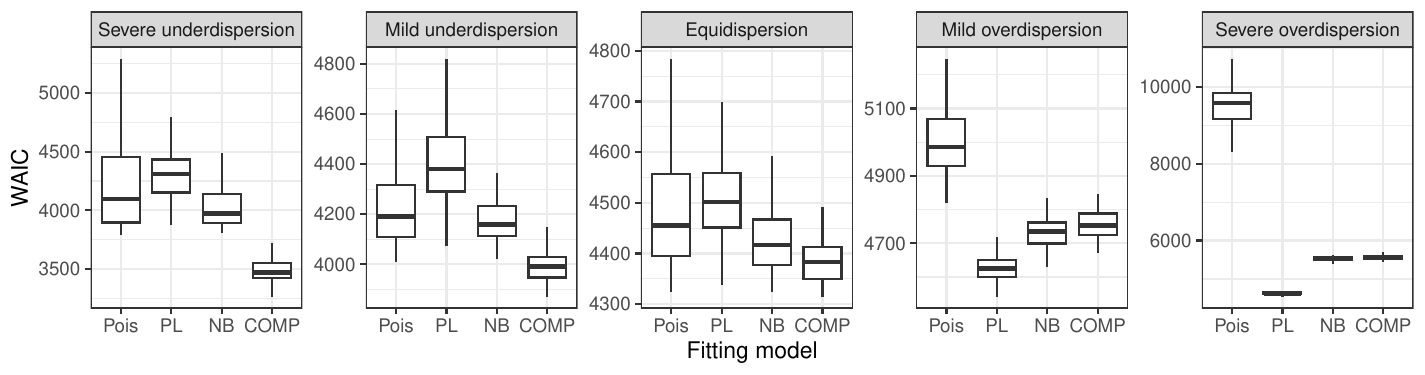}
    \caption{The box plots for WAIC values, across 100 simulations, for four different models (x-axis) for each simulation scenario (each column).}
    \label{sup:fig:waicall}
\end{figure}

Figure~\ref{sup:fig:waicall} shows box plots for WAIC values, across 100 simulations, for the Poisson, PL, NB, and COMP$_{\mu}$ models for each simulation scenario. Smaller WAIC is better.

\newpage
\subsection{Spatial zero-inflated count data}

Table~\ref{sup:tab:simszicomp} shows estimated posterior medians and 95\% HPD intervals for model parameters for underdispersed and overdispersed datasets. We observe that all of their 95\% HPD intervals cover the true values.

\begin{table}[H]
\caption{(a) True values, estimated posterior medians, and 95\% highest posterior density (HPD) intervals for $\beta_{11}$, $\beta_{12}$, $\beta_{21}$, $\beta_{22}$, and $\log(\nu)$ for a under-dispersed data set. (b) Those for a over-dispersed data set. The estimates are close to the true values. All of their 95\% HPD intervals cover the true values.}
\label{sup:tab:simszicomp}
\centering
\begin{tabular}{lrrr c rrr}
    \toprule
     & \multicolumn{3}{c}{Under-dispersed data} && \multicolumn{3}{c}{Over-dispersed data}\\
     \cmidrule{2-4} \cmidrule{6-8}
     & Truth & Median & 95\% HPD && Truth & Median & 95\% HPD \\\midrule
    $\beta_{11}$ & -2.00 & -2.64 & (-3.16, -1.91) && -2.00 & -2.34 & (-2.79,-1.86)\\
    $\beta_{12}$ & 1.00 & 1.61 & (0.89, 2.10) && 1.00 & 1.31 & (0.86, 1.75)\\
    $\beta_{21}$ & 2.00 & 1.95 & (1.85, 2.32) && 2.00 & 1.91 & (1.77, 2.09)\\
    $\beta_{22}$ & 2.00 & 2.02 & (1.78, 2.21) && 2.00 & 2.10 & (2.00, 2.23)\\
    $\log(\nu)$ & 0.53 & 0.47 & (-0.82, 0.62) && -0.36 & -0.37 & (-0.56, -0.27)\\\bottomrule
\end{tabular}
\end{table}

\begin{figure}[H]
\centering
\includegraphics[width = 0.75\textwidth]{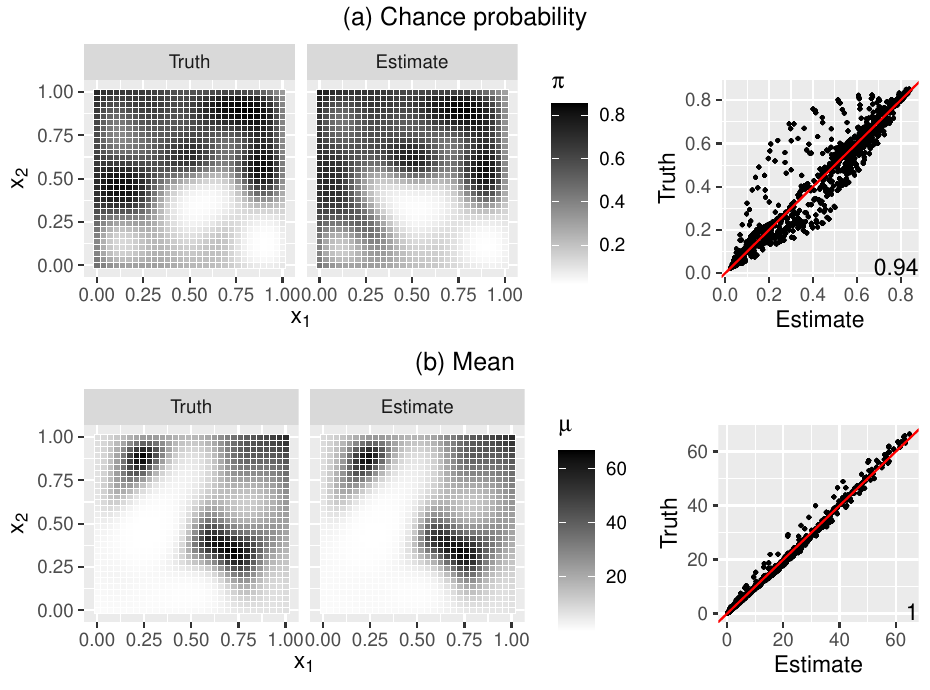}
\caption{(a) Maps of true and estimated chance probabilities $\pi_i$ and a scatter plot with the estimated correlation coefficient in the bottom right. (b) Those for means $\mu_i$. The true spatial distributions are recovered very well. \label{sup:fig:szicompOverMap}}
\end{figure}

Figure~\ref{sup:fig:szicompOverMap} shows estimated spatial patterns closely mirror the true spatial patterns for the overdispersed data. This indicates that our approach recovers well the underlying spatial patterns in the data.

\begin{figure}[H]
    \centering
    \includegraphics[width = 0.75\textwidth]{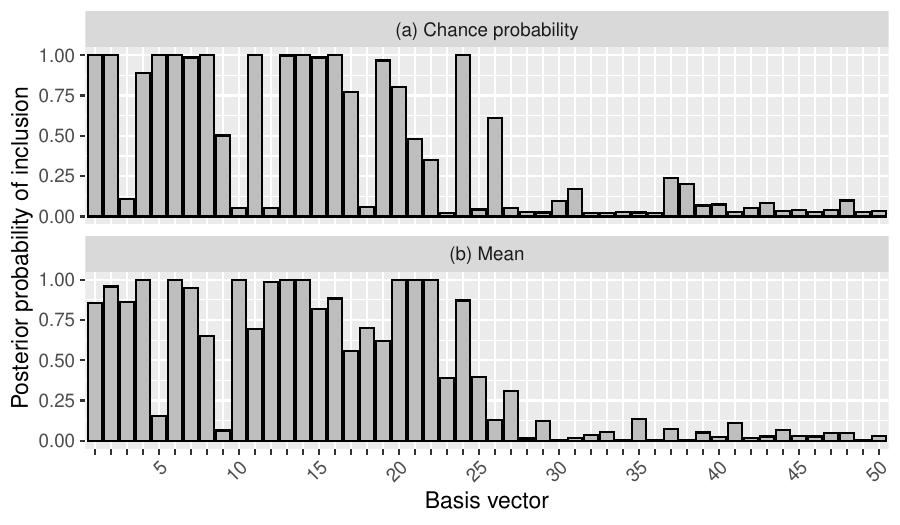}
    \caption{Estimated posterior probabilities of including basis vectors for the chance probability (a) and the mean (b) for a simulated zero-inflated underdispersed dataset.}
    \label{sup:fig:szicompUnderBasis}
\end{figure}
\begin{figure}[H]
    \centering
    \includegraphics[width = 0.75\textwidth]{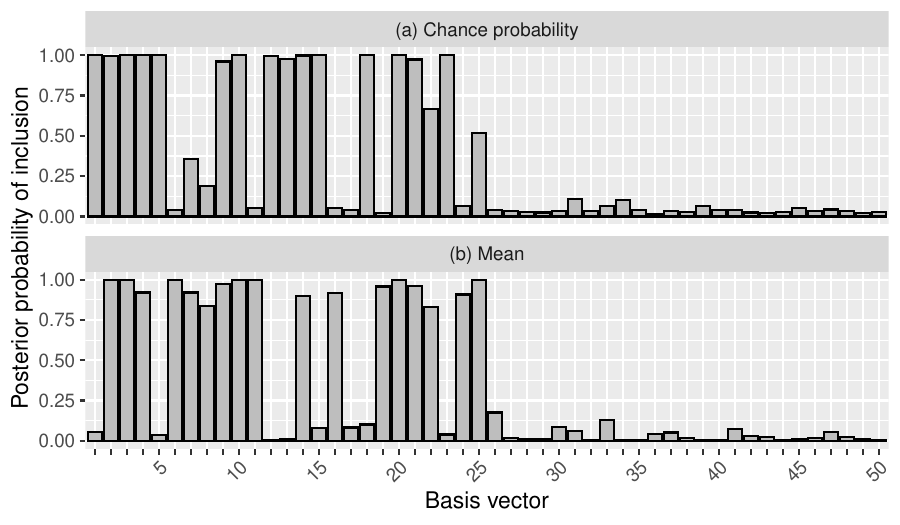}
    \caption{Estimated posterior probabilities of including basis vectors for the chance probability (a) and the mean (b) for a simulated zero-inflated underdispersed dataset.}
    \label{sup:fig:szicompOverBasis}
\end{figure}

Figures \ref{sup:fig:szicompUnderBasis} and \ref{sup:fig:szicompOverBasis} shows posterior probabilities for basis-vector selection for a zero-inflated underdispersed dataset and a zero-inflated overdispersed dataset, respectively. We observe that all significant basis vectors (who have quite large magnitudes of true basis coefficients) have large posterior probabilities of inclusion. 

\begin{figure}[H]
\centering
\includegraphics[width = 0.8\textwidth]{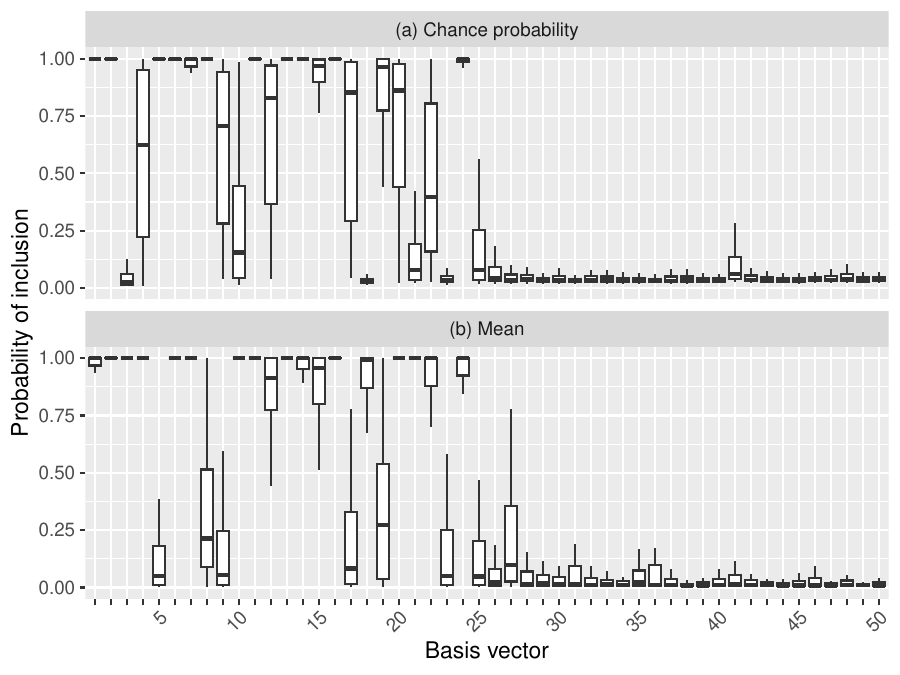}
\caption{Box plots for estimated posterior probabilities for basis-vector selection, across 100 simulations, for the chance probability (a) and the mean (b) for a simulated underdispersed zero-inflated count dataset.\label{fig:szicompUnderBasisBox}}
\end{figure}
\begin{figure}[H]
\centering
\includegraphics[width = 0.8\textwidth]{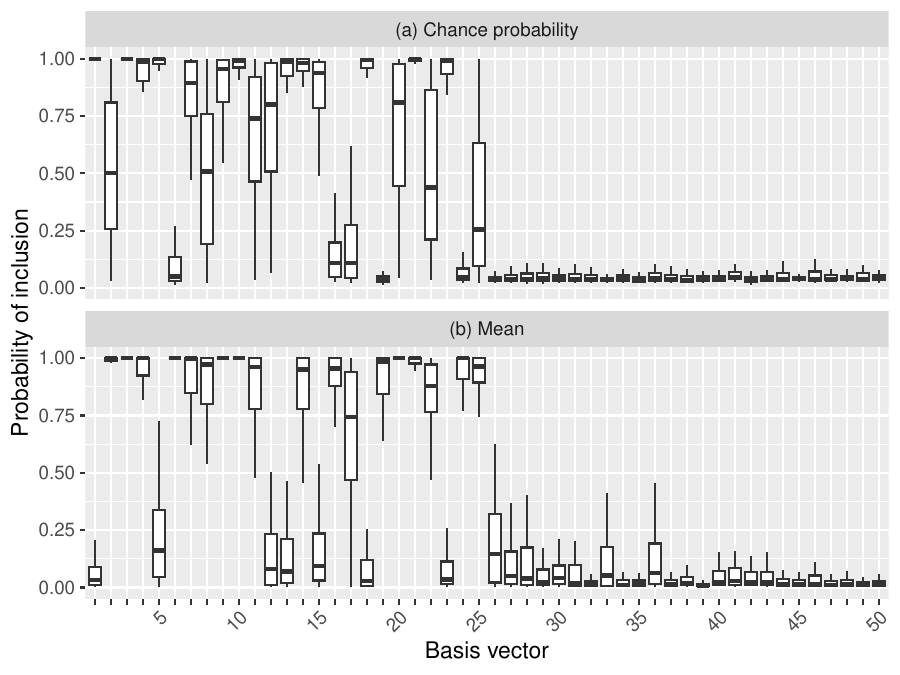}
\caption{Box plots for estimated posterior probabilities for basis-vector selection, across 100 simulations, for the chance probability (a) and the mean (b) for a simulated overdispersed zero-inflated count dataset.\label{fig:szicompOverBasisBox}}
\end{figure}

We examine the distributions of estimated posterior means for each $I_{\gamma_j}$ and $I_{\delta_j}$ across the 100 samples.
Figures \ref{fig:szicompUnderBasisBox} and \ref{fig:szicompOverBasisBox} show the results for the underdispersed and overdispersed datasets, respectively. We find that all significant basis vectors (whose box plots are close to 1) have quite large magnitudes of true basis coefficients. The insignificant ones (whose box plots are close to 0) are found to have true basis coefficient values either close to 0 or equal to 0. This implies that our methodology can successfully detect important basis vectors for both chance probability and mean.

\section{Supplemental figures and tables for Section 5}

Here we provide supplemental figures and tables for the real data analysis described in Section 5 of the manuscript.

\subsection{Texas HPV cancer data}

\begin{table}[H]
    \caption{Description of the covariate variables of HPV-related cancer}
    \label{sup:tab:hpv}
    \centering
        \begin{tabular}{ll}
        \toprule
        Variable & Description\\ \midrule
        Male & Proportion of males\\
        Female & Proportion of females (reference level)\\
        Age 0--14 & Proportion of people aged 0 to 19\\
        Age 15--19 & Proportion of people aged 15 to 19\\
        Age 20--29 & Proportion of people aged 20 to 29\\
        Age 30+ & Proportion of people aged 30 years or older (reference level)\\
        White & Proportion of White (reference level)\\
        Black & Proportion of Black or African American\\
        Other & Proportion of others\\
        Chlamydia & Crude chlamydia incidence rate per 100,000 people\\
        Smoking & \makecell[l]{Proportion of people aged 18 years or older who smoke\\at least 100 cigarettes}\\
        Uninsured & \makecell[l]{Proportion of people under 64 years of age without\\health insurance}\\
        Income & Per capita income in the past 12 months\\
        Poverty & Proportion of people in poverty\\
        Urban population & \makecell[l]{Proportion of the total population of the county\\represented by the urban population}\\
        \bottomrule
    \end{tabular}
\end{table}

Table~\ref{sup:tab:hpv} provides description of the covariate variables for the Texas HPV cancer data.

\begin{figure}[H]
\centering
\includegraphics[width = 0.6\textwidth]{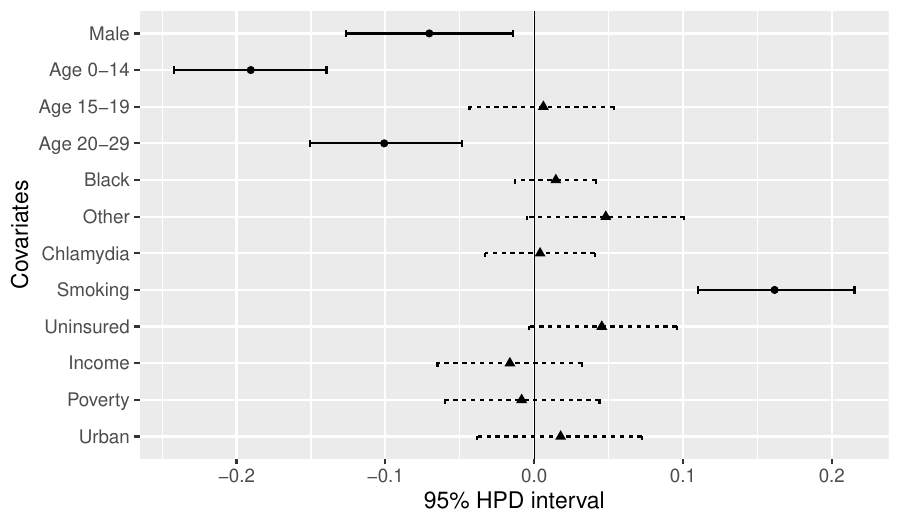}
\caption{Estimated posterior medians (shaded dots or triangles) and 95\% highest posterior density (HPD) intervals (horizontal solid or dashed bars) for covariate coefficients. The shaded dot and horizontal solid bar represent that the HPD interval does not include zero. The triangle and horizontal dashed bar represent that the HPD interval includes zero. \label{sup:fig:hpvcoef}}
\end{figure}

Figure~\ref{sup:fig:hpvcoef} shows the estimated posterior median and 95\% highest posterior density (HPD) interval for each of the covariate coefficients. We have approximately 0.23, 0.29, and 0.21 for the acceptance probabilities of $\bs{\beta}$, $\nu$, and $\bs{\gamma}$, respectively.

\begin{figure}[H]
\centering
\includegraphics[width = 0.6\textwidth]{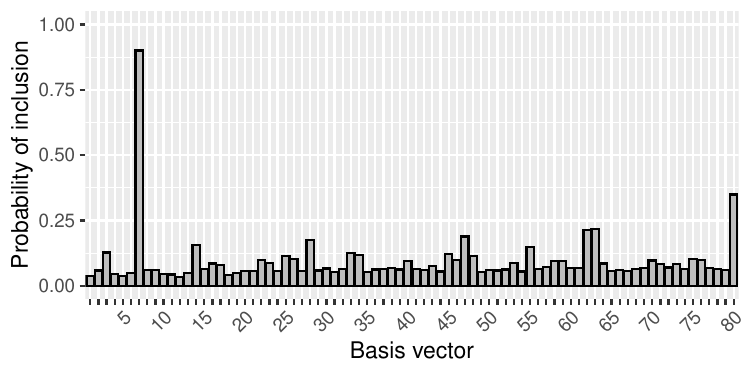}
\caption{The estimated posterior probabilities of including basis vectors. One basis vector is found to be a significant predictor. \label{sup:fig:hpvbasis}}
\end{figure}

Figure~\ref{sup:fig:hpvbasis} displays the estimated posterior probability for each of the basis vectors we consider in this example. One basis vector is found to be a significant predictor. The acceptance probability for the basis-vector selection is approximately 0.11 on average.

\begin{figure}[H]
\centering
\includegraphics[width = \textwidth]{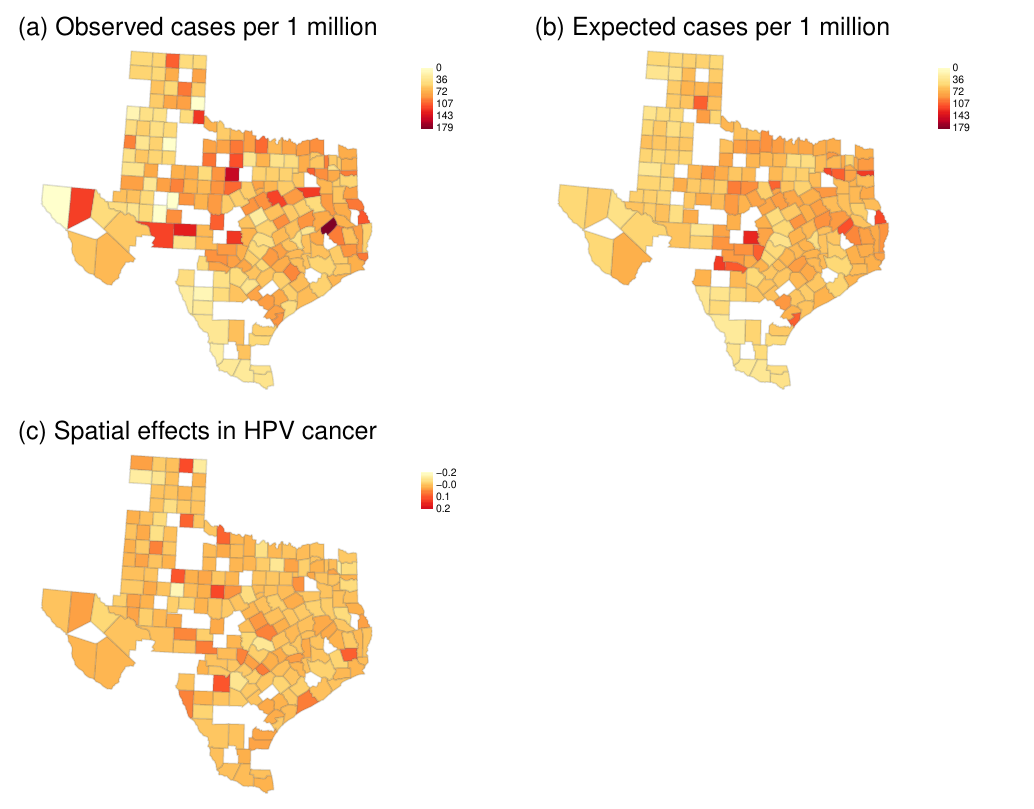}
\caption{(a) The observed number of HPV-related cancers per 1 million people. (b) The model estimate of the expected number of HPV-related cancers per 1 million people. (c) The model estimates of the spatial random effects. \label{sup:fig:hpvmap}}
\end{figure}

Figure~\ref{sup:fig:hpvmap} (a) represents the observed number of HPV-related cancer cases per 1 million people for each county during the period of 2006--2012 in Texas. Figure~\ref{sup:fig:hpvmap} (b) shows the model estimate of the expected number of HPV-related cancer cases per 1 million people for each county.  Figure~\ref{sup:fig:hpvmap} (c) displays the model estimates of the spatial random effects $\bs{u} \approx \textbf{B} \bs{\delta}$.

\subsection{US vaccine refusal data}

\setlength{\tabcolsep}{3pt}
\begin{table}[H]
    \caption{Description of the covariate variables of vaccine refusal}
    \label{sup:tab:vaccine}
    \centering
        \begin{tabular}{lll}
        \toprule
        \multicolumn{2}{l}{Variable} & Description\\\midrule
        \multicolumn{3}{l}{\textit{Measurement variables}}\\\cmidrule{2-3}
         & Physician-patient interactions & Number of physician-patient interactions \\
         & Health insurance & Proportion of people with health insurance\\
         & Pediatrician reporting & \makecell[l]{Rate at which a pediatrician voluntarily reports\\non-billable diagnoses}\\ \midrule
        \multicolumn{3}{l}{\textit{Demographic or socioeconomic variables}}\\\cmidrule{2-3}
         & Household size & \makecell[l]{Average number of individuals living\\in a single household}\\
        & Religious congregations & \makecell[l]{Per capita number of congregations of religions\\historically opposed to vaccination}\\
         & Limited English proficiency & Proportion of people who are not proficient in English\\
         & Private school & Proportion of children who attend private school\\
         & High income & \makecell[l]{Proportion of people in the upper 20\% quantile\\of income in the US}\\
         & Same area & \makecell[l]{Proportion of people living in the same county\\one year prior}\\
         & State law leniency & Exemption law effectiveness index\\
         & State autism & \makecell[l]{Among families with more than 1 child, the proportion\\with a current or past diagnosis of autism}\\\bottomrule
    \end{tabular}
\end{table}

Table~\ref{sup:tab:vaccine} provides description of the covariate variables for the US vaccine refusal data.

\begin{figure}[H]
\centering
\includegraphics[width = \textwidth]{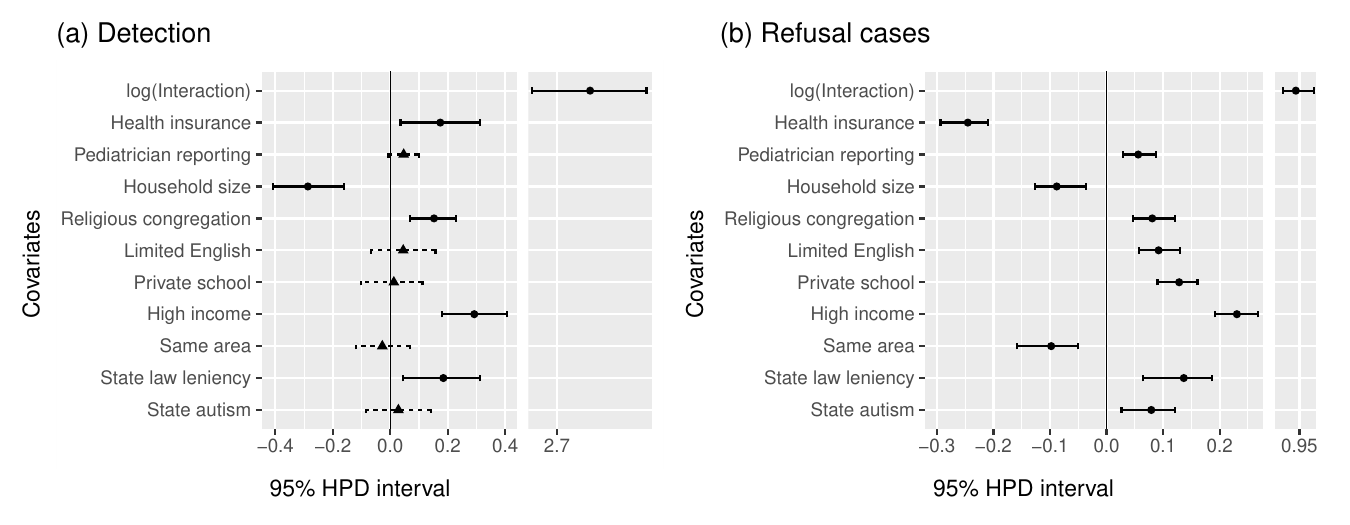}
\caption{(a) Estimated posterior medians (shaded dots or triangles) and 95\% highest posterior density (HPD) intervals (horizontal solid or dashed bars) for covariate coefficients for the detection of refusal. (b) Those for vaccine refusal cases. The shaded dot and horizontal solid bar represent that the HPD interval does not include zero. The triangle and horizontal dashed bar represent that the HPD interval includes zero. \label{fig:refusecoef}}
\end{figure}

Figure~\ref{fig:refusecoef} (a) shows the estimated posterior median and 95\% highest posterior density (HPD) interval for each of the covariate coefficients for the detection of refusal. Figure~\ref{fig:refusecoef} (b) displays the estimated posterior median and 95\% highest posterior density (HPD) interval for each of the covariate coefficients for the vaccine refusal cases. We have approximately 0.23, 0.21, 0.26, 0.20, and 0.24 for the acceptance probabilities of $\bs{\beta}_1$, $\bs{\beta}_2$, $\nu$, $\bs{\gamma}$, and $\bs{\gamma}$, respectively. 

\begin{figure}[H]
\centering
\includegraphics[width = 0.8\textwidth]{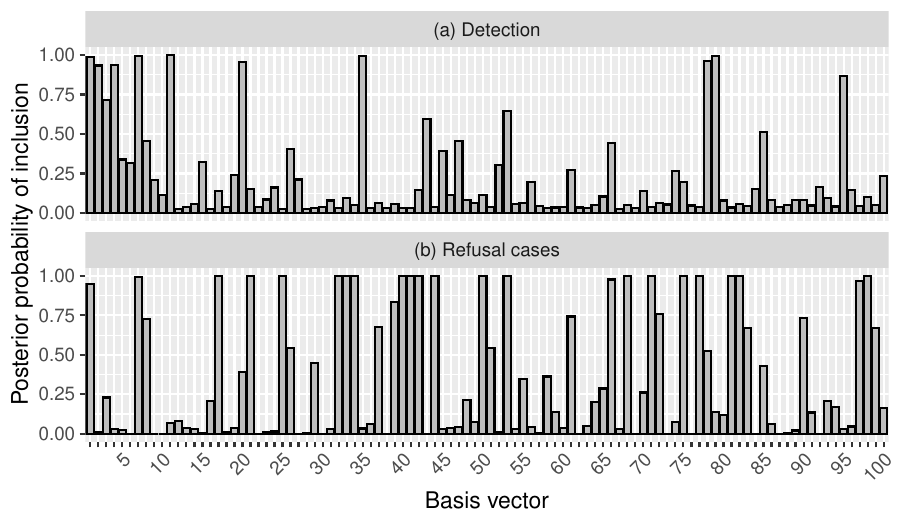}
\caption{(a) The estimated posterior probabilities of including basis vectors for detection of refusal. (b) Those for vaccine refusal cases. \label{sup:fig:refuseBasis}}
\end{figure}

Figure~\ref{sup:fig:refuseBasis} shows the estimated posterior probabilities for each of the basis vectors we consider in this example. 
For the detection of refusal, the acceptance probability for the basis-vector selection is approximately 0.086 on average. For the vaccine refusal cases, the acceptance probability for the basis-vector selection is approximately 0.016 on average.

\begin{figure}[H]
\centering
\includegraphics[width = 0.9\textwidth]{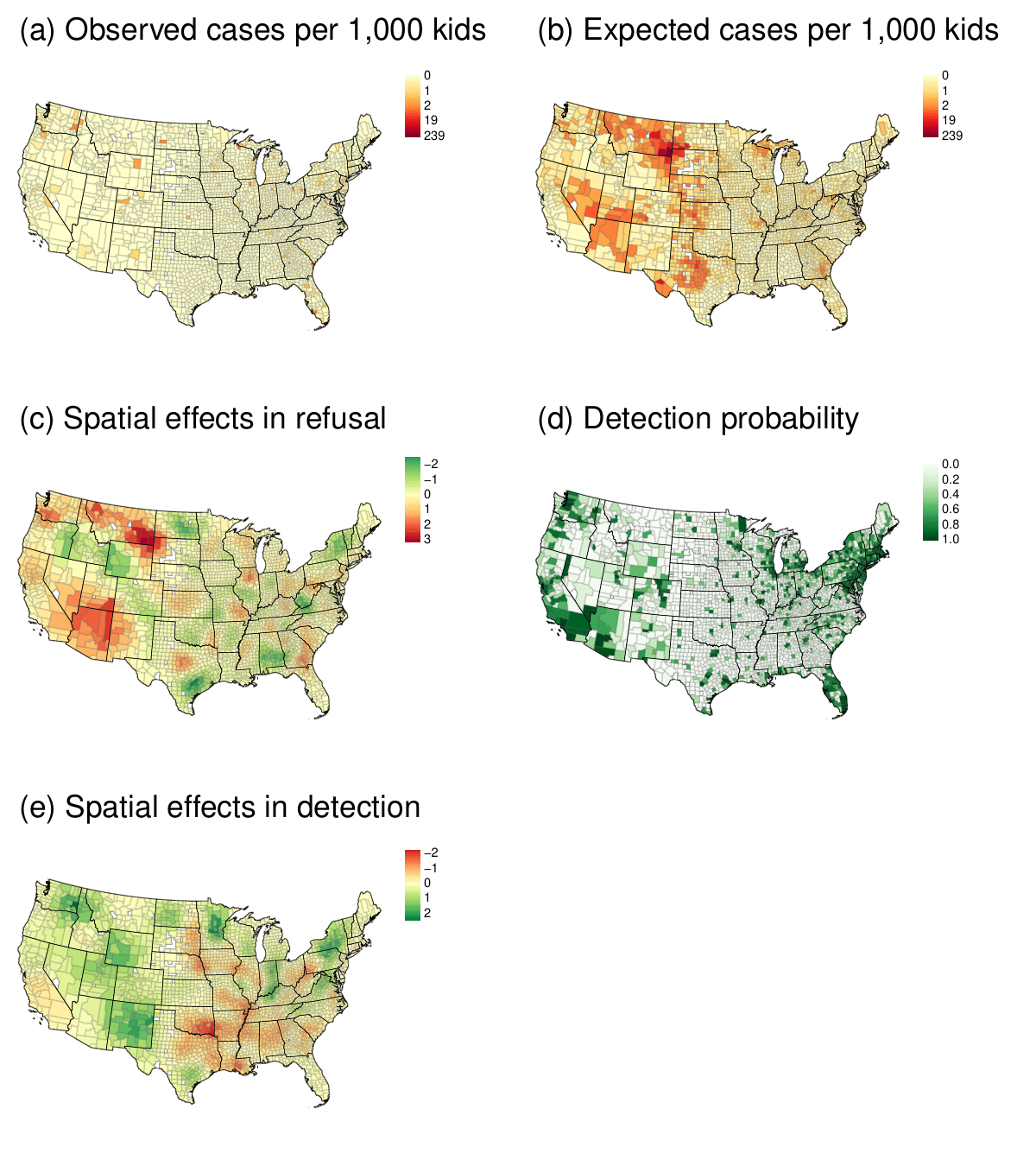}
\caption{(a) The observed number of refusal cases per 1,000 kids. (b) The model estimates of the expected number of refusal cases per 1,000 kids. (c) The model estimates of the spatial random effects for refusal cases. (d) The model estimates of the detection probabilities. (e) The model estimates of the spatial random effects for detection.\label{sup:fig:refusemap}}
\end{figure}

Figure~\ref{sup:fig:refusemap} (a) represents the observed number of vaccine refusal cases per 1,000 kids for each county in June 2015 in the United States. Figure~\ref{sup:fig:refusemap} (b) shows the model estimate of the expected number of vaccine refusal cases per 1,000 kids for each county. Figure~\ref{sup:fig:refusemap} (c) displays the model estimates of the spatial random effects $\bs{u} \approx \textbf{B} \bs{\delta}$ for refusal cases. Figure~\ref{sup:fig:refusemap} (d) shows the model estimate of the detection probability of refusal for each county. Figure~\ref{sup:fig:refusemap} (e) represents the model estimates of the spatial random effects $\bs{u} \approx \textbf{B} \bs{\gamma}$ for the detection of refusal. Similar results are observed for the other months.

\bibliographystyle{apalike} 
\bibliography{refs}